\documentclass[a4paper,11pt]{article}
\usepackage[a4paper,margin=25mm]{geometry}
\usepackage{microtype}
\usepackage{amsmath,amssymb,mathtools,bm}
\usepackage{graphicx,booktabs,longtable,array}
\usepackage{xcolor}
\definecolor{linkblue}{RGB}{0,50,232}
\definecolor{equationpink}{RGB}{232,0,120}
\usepackage[colorlinks=true,
 linkcolor=linkblue,
 citecolor=linkblue,
 urlcolor=linkblue]{hyperref}
\hypersetup{
 pdftitle={Holographic Renormalization of String-Derived Lovelock--Horndeski Theory},
 pdfauthor={Tianhao Wu},
 pdfsubject={Holographic renormalization in Lovelock--Horndeski gravity}
}
\makeatletter
\renewcommand{\eqref}[1]{%
 \textup{\hyperref[#1]{\textcolor{equationpink}{\tagform@{\ref*{#1}}}}}%
}
\makeatother
\numberwithin{equation}{section}
\allowdisplaybreaks
\renewcommand{\Box}{\mathord{\vcenter{\hrule height .08em
 \hbox{\vrule width .08em height .65em\kern .58em
 \vrule width .08em height .65em}\hrule height .08em}}}

\newcommand{\dd}{\mathrm d}
\newcommand{\tr}{\operatorname{tr}}
\newcommand{\rank}{\operatorname{rank}}

\newcommand{\cA}{\mathcal A}
\newcommand{\cB}{\mathcal B}
\newcommand{\cC}{\mathcal C}
\newcommand{\cE}{\mathcal E}
\newcommand{\cG}{\mathcal G}
\newcommand{\cH}{\mathcal H}
\newcommand{\cI}{\mathcal I}
\newcommand{\cJ}{\mathcal J}
\newcommand{\cL}{\mathcal L}
\newcommand{\cO}{\mathcal O}
\newcommand{\cP}{\mathcal P}
\newcommand{\cQ}{\mathcal Q}
\newcommand{\cS}{\mathcal S}

\newcommand{\cU}{\mathcal U}
\newcommand{\cV}{\mathcal V}

\DeclareMathAlphabet{\mathsf}{OT1}{cmr}{m}{n}

\title{Holographic Renormalization of String-Derived Lovelock--Horndeski Theory}
\author{Tianhao Wu\\[0.35em]
\normalsize Department of Physics, University of Illinois Urbana--Champaign,\\
\normalsize Urbana, Illinois 61801, USA\\[0.25em]
\normalsize\href{mailto:twu49@illinois.edu}{twu49@illinois.edu}}
\date{}

\begin{document}
\maketitle

\begin{abstract}
String-derived higher-curvature scalar--tensor gravities encode microscopic coupling data in boundary response, raising the question of whether holographic observables can reconstruct the underlying higher-dimensional parameters. We answer this question for the five-dimensional string-derived Lovelock--Horndeski (SDLH) theory on its exact linear-dilaton asymptotically locally AdS branch, constructing the renormalized generating functional for an arbitrary boundary metric and spacetime-dependent scalar source. A boundary-covariant radial hierarchy unifies the variational problem, local backreaction, logarithmic obstruction, finite one-point functions, and Ward identities. Two response determinants organize the recursion, resonant obstructions, and metric--scalar mixing. The Weyl anomaly condenses into an Euler density, a Weyl-squared density, and a single curvature--scalar square whose paired variations generate the metric and scalar obstructions. The resulting renormalized functional carries string-selected coupling data into boundary geometry, operator response, anomaly coefficients, and a calculable interface with gravitational observables. On the regular branch, four scalar-normalization-invariant holographic combinations admit a global rational inverse to the continuous reduced couplings. At fixed compactification dimension the map has maximal rank, while the curvature-anomaly sum reconstructs the higher-dimensional Gauss--Bonnet coefficient without sign ambiguity. Holographic response thus provides an explicit, overdetermined boundary fingerprint of the underlying string reduction.
\end{abstract}

\section{Introduction}
\label{sec:introduction}

The AdS/CFT correspondence turns radial gravitational dynamics into a source-resolved description of quantum field theory. After Euclidean continuation, the renormalized on-shell action for a bulk metric \(G_{MN}\) and a scalar field \(\phi\) supplies the generating functional through
\begin{equation}
 Z_{\rm grav}\!\left[g_{(0)},\phi_{(0)}\right]
 =\exp\!\left[-S_{\rm ren}\!\left[g_{(0)},\phi_{(0)}\right]\right]
 =Z_{\rm QFT}\!\left[g_{(0)},\phi_{(0)}\right],
 \label{eq:intro-gkpw}
\end{equation}
where \(g_{(0)ij}(x)\) and \(\phi_{(0)}(x)\) act as the sources for the stress tensor and the scalar operator. The foundational AdS/CFT constructions established this relation as the bridge between bulk boundary data and quantum-field-theory generating functionals~\cite{Maldacena1998,GubserKlebanovPolyakov1998,Witten1998,AharonyEtAl2000}. To evaluate this relation, one solves the near-boundary equations, completes the variational problem, and adds the counterterm action. Functional differentiation then yields finite one-point functions; radial diffeomorphism covariance yields the boundary Ward identities. This sequence connects bulk couplings directly to local response data.

Fefferman--Graham geometry provides the natural language for this construction. Near a four-dimensional conformal boundary, the radial expansion resolves the bulk equations by dilatation weight and converts curvature into covariant tensors of the source metric~\cite{FeffermanGraham1985,HenningsonSkenderis1998}. Holographic renormalization organizes the divergent action, its local counterterms, and the finite canonical momenta in the same expansion~\cite{BalasubramanianKraus1999,deHaroSkenderisSolodukhin2001,BianchiFreedmanSkenderis2002,Skenderis2002}. The Hamilton--Jacobi formulation expresses the radial hierarchy as a derivative descent for the on-shell functional~\cite{deBoerVerlindeVerlinde2000,MartelliMueck2003,Papadimitriou2010,Papadimitriou2011,ElvangHadjiantonis2016}. Local scalar sources introduce metric--scalar mixing at every active derivative weight. The momentum constraint measures local exchange between the stress tensor and the scalar operator, and the logarithmic coefficient encodes the generalized Weyl anomaly.

Higher-curvature scalar--tensor theories enrich this dictionary through curvature response, operator mixing, and completed canonical momenta. Lovelock densities preserve second-order metric equations in their natural dimensions~\cite{Lovelock1971}; the Gauss--Bonnet density also arises as the leading curvature-squared string correction~\cite{Zwiebach1985,BoulwareDeser1985,Wheeler1986}. Horndeski interactions and covariant Galileons supply the corresponding second-order scalar--tensor structures~\cite{Horndeski1974,NicolisRattazziTrincherini2009,DeffayetEspositoFareseVikman2009,KobayashiYamaguchiYokoyama2011,CharmousisEtAl2012,KovacsReall2020}. Here ``Lovelock--Horndeski'' denotes the five-dimensional Lovelock--Galileon subset whose metric and scalar equations retain second differential order. Their boundary variations generate characteristic canonical currents and Dirichlet completions~\cite{Myers1987,DyerHinterbichler2009,PadillaSivanesan2012,HarlowWu2020}. These sectors meet naturally in dimensional reductions of string effective actions, where field-redefinition freedom organizes an equivalence class of higher-derivative coefficient frames~\cite{MetsaevTseytlin1987,GrossSloan1987,BergshoeffdeRoo1989,ChemissanydeRooPanda2007,DuffNilssonPope1986}. The reduced couplings consequently retain a calculable memory of the higher-dimensional action, the reduction ansatz, and the internal geometry.

Higher-curvature anomaly calculations extract Euler and Weyl coefficients from the logarithmic on-shell action~\cite{HenningsonSkenderis1998,DeserSchwimmer1993,Duff1994,ImbimboEtAl2000,NojiriOdintsov2000}; scalar-inclusive renormalization develops local counterterms and operator responses~\cite{deHaroSkenderisSolodukhin2001,Skenderis2002,KanitscheiderSkenderisTaylor2008,Papadimitriou2011,JahnkeMisobuchiTrancanelli2015}; and Horndeski holography exhibits linear scalar hair, critical response, and modified RG data~\cite{AnabalonCisternaOliva2014,LiLu2018,LiLuZhang2019,CharmousisGouterauxKiritsis2012,CaceresEtAl2024}. The five-dimensional string-derived Lovelock--Horndeski (SDLH) theory introduced by Wu and Stone in~\cite{WuStone2026} provides a setting in which higher-curvature and derivative scalar interactions support a linear-dilaton AdS branch. We construct its renormalized generating functional for an arbitrary boundary metric and spacetime-dependent scalar source.

The five-dimensional SDLH action takes the form
\begin{equation}
 \begin{aligned}
 S_{\rm bulk}=\frac{1}{16\pi G_5}\int \dd^5x\sqrt{-G}\,
 \bigg( &R+\alpha_0X+\alpha_1\cL_{\rm GB}
 +\alpha_2\cG^{MN}\nabla_M\phi\nabla_N\phi +\alpha_3X\Box\phi+\alpha_4X^2\bigg) ,
 \end{aligned}
 \label{eq:intro-action}
\end{equation}
where \(X_M\equiv\nabla_M\phi,\; X\equiv X_MX^M\). \(\cG_{MN}\) is the five-dimensional Einstein tensor and
\(\cL_{\rm GB}=R_{MNRS}R^{MNRS}-4R_{MN}R^{MN}+R^2\). We normalize the Einstein--Hilbert coefficient to unity and retain \(\alpha_0\) as the scalar normalization. The coupled metric--scalar branch equations determine the AdS scale, while the heterotic construction and Kaluza--Klein reduction fix the SDLH coupling trajectory. The string construction, linear-dilaton branch, covariant anomaly, and holographic RG structure were established in~\cite{WuStone2026}.

We retain the exact rational dependence of the radial response on
\((\alpha_0,\ldots,\alpha_4,\ell,s)\). Allowing local metric and scalar
sources thereby promotes the string-selected coupling trajectory to a source-resolved
map into boundary geometry, operator mixing, and momentum exchange.

The relevant asymptotic solution carries a logarithmic radial scalar profile. In Fefferman--Graham gauge we use
\begin{equation}
 \begin{aligned}
 \dd s^2&=\frac{\ell^2}{4\rho^2}\dd\rho^2
 +\rho^{-1}g_{ij}(\rho,x)\dd x^i\dd x^j,\\
 g_{ij}&=g_{(0)ij}+\rho g_{(2)ij}
 +\rho^2\!\left(g_{(4)ij}^{\rm tot}+\log\rho\,h_{(4)ij}\right)+\cdots,\\
 \phi&=s\log\rho+\phi_{(0)}+\rho\phi_{(2)}
 +\rho^2\!\left(\phi_{(4)}^{\rm tot}+\log\rho\,\psi_{(4)}\right)+\cdots .
 \end{aligned}
 \label{eq:intro-fg}
\end{equation}
We use a dimensionless FG coordinate; restoring a renormalization scale replaces \(\log\rho\) by \(\log(\rho\mu^2)\). The radial slope \(s\) supports the leading AdS geometry. A smooth \(\phi_{(0)}(x)\) enters the bulk stress tensor at derivative weight two, since
\(G^{ij}\partial_i\phi\partial_j\phi=\rho g_{(0)}^{ij}\partial_i\phi_{(0)}\partial_j\phi_{(0)}+O(\rho^2)\), and therefore preserves the leading asymptotically locally AdS (AlAdS) geometry. The source pair \((g_{(0)},\phi_{(0)})\) probes the full local operator response: curvature activates the tensor channel, scalar gradients activate metric--scalar mixing, and the four \(E_{\rho i}\) equations become genuine momentum-exchange constraints. The regular response manifold is defined by \(s\Delta_T\Delta_S\ne0\). On this manifold the radial equations furnish invertible tensor and scalar response channels with determinants \(\Delta_T\) and \(\Delta_S\).

We begin with the complete radial variational problem. Writing \(\gamma_{ij}=\rho^{-1}g_{ij}\), \(K_{ij}=\frac12\cL_n\gamma_{ij}\), \(v=n^M\nabla_M\phi\), and \(X_i=D_i\phi\), the outward-normal Dirichlet surface density is
\begin{equation}
 \cB_D=2K
 +4\alpha_1\!\left(J-2\widehat{\cG}_{ij}K^{ij}\right)
 +\alpha_2\!\left(K^{ij}X_iX_j-KX_kX^k\right)
 -\alpha_3v\!\left(X_kX^k+\frac13v^2\right).
 \label{eq:intro-dirichlet}
\end{equation}
Here \(\widehat{\cG}_{ij}\) is the Einstein tensor of the cutoff metric and \(J\) is the cubic extrinsic-curvature scalar entering the Myers completion. In Eq.~\eqref{eq:intro-dirichlet}, we combine the Einstein/Gibbons--Hawking--York, Gauss--Bonnet/Myers, Einstein-tensor kinetic, and cubic Galileon sectors in one source-variation convention~\cite{GibbonsHawking1977,Myers1987,Davis2003,PadillaSivanesan2012}. This produces completed metric and scalar canonical currents whose divergent and finite coefficients follow directly from Eq.~\eqref{eq:intro-fg}. The radial construction relates the constraint equations, counterterm descent, and observable variations within the same source convention.

We solve the coupled Fefferman--Graham recursion at exact \((\alpha_0,\ldots,\alpha_4,\ell,s)\). At weight two, boundary covariance, parity, and shift symmetry close the solution on the nine coefficients of the tensor--scalar basis, all fixed locally by \((g_{(0)},\phi_{(0)})\). At weight four, we separate the source-local particular solution, logarithmic obstruction, and normalizable metric--scalar response. Accordingly, \(g_{(4)}^{\rm tot}=g_{(4)}^{\rm local}+g_{(4)}^{\rm state}\) and \(\phi_{(4)}^{\rm tot}=\phi_{(4)}^{\rm local}+\phi_{(4)}^{\rm state}\), while the exact determinants \(\Delta_T\) and \(\Delta_S\) organize the tensor and scalar response channels.

We obtain the logarithmic divergence as a compact covariant functional. Defining
\begin{equation}
 \Xi\equiv X_kX^k+2s\Box_{(0)}\phi_{(0)}
 -\frac23s^2R[g_{(0)}],
 \label{eq:intro-Q}
\end{equation}
we obtain
\begin{equation}
 \cA_{\rm reg}
 =a_EE_4+a_CC_{ijkl}C^{ijkl}-\frac{\kappa}{4}\Xi^2,
 \label{eq:intro-anomaly}
\end{equation}
with exact rational functions \(a_E\), \(a_C\), and \(\kappa\) of the SDLH couplings and branch data. Expanding Eq.~\eqref{eq:intro-anomaly} generates the nine-density curvature--scalar basis. Its functional variations span an eight-dimensional response space, with the four-dimensional Euler/Lanczos density furnishing the topological null direction. The negative of Eq.~\eqref{eq:intro-anomaly} supplies the logarithmic counterterm density. The paired metric and scalar variations, followed by inversion of the response matrix, determine \(h_{(4)ij}\) and \(\psi_{(4)}\) and connect the anomaly directly to the logarithmic Fefferman--Graham obstruction.

The finite completed currents define
\begin{equation}
 \delta S_{\rm ren}=\int \dd^4x\sqrt{-g_{(0)}}
 \left(\frac12\langle T^{ij}\rangle\delta g_{(0)ij}
 +\langle\cO_\phi\rangle\delta\phi_{(0)}\right).
 \label{eq:intro-onepoint-definition}
\end{equation}
The finite completed currents give the stress tensor and scalar one-point function as exact covariant coefficient operations. Their response-bearing part depends on the total weight-four coefficients,
\begin{equation}
 g_{(4)}^{\rm tot}=g_{(4)}^{\rm local}[g_{(0)},\phi_{(0)}]+g_{(4)}^{\rm state},
 \qquad
 \phi_{(4)}^{\rm tot}=\phi_{(4)}^{\rm local}[g_{(0)},\phi_{(0)}]+\phi_{(4)}^{\rm state}.
 \label{eq:intro-weight-four-split}
\end{equation}
The local terms are fixed by the near-boundary recursion, while the normalizable terms encode the quantum state selected by the interior bulk solution. The homogeneous reduction reproduces the exact metric--scalar mixing coefficients and the normalizable constraint
\begin{equation}
 8\alpha_0s\,\phi_{(4)}^{\rm state}
 +\left(\alpha_0s^2-3\right)t_4=0,
 \qquad t_4\equiv\tr g_{(4)}^{\rm state}.
 \label{eq:intro-normalizable}
\end{equation}

We derive both Ward identities from radial covariance. Boundary functional covariance and the radial projection of the bulk diffeomorphism Noether identity give
\begin{equation}
 D_i\langle T^i{}_j\rangle
 -\langle\cO_\phi\rangle D_j\phi_{(0)}
 =\frac{\overline\cE_{\rho j}^{\rm nonlog}
 -\tfrac12\overline\cE_{\rho j}^{\log}}
 {4\pi G_5\ell}=0.
 \label{eq:intro-diffeo-ward}
\end{equation}
The radial dilatation equation gives the generalized trace identity in the minimal finite scheme
\begin{equation}
 \langle T^i{}_i\rangle-2s\langle\cO_\phi\rangle
 =-\frac{\cA_{\rm reg}}{8\pi G_5}.
 \label{eq:intro-trace-ward}
\end{equation}
The first identity measures the momentum transferred from the local scalar source to the stress tensor. The second identifies \(2s\) as the beta-spurion weight generated by the logarithmic radial scalar profile. Their common Noether origin ties the state response, counterterms, anomaly, and finite canonical momenta to the same renormalized functional.

We use the exact coupling dependence to define a coefficient-frame response map. Choose a compactification and coefficient frame \(\Theta_{\rm str}\), and use the regular-branch chart \(\cI_A=(\alpha_0,\alpha_2,\alpha_3;\ell,s)\), with \(\alpha_1\) and \(\alpha_4\) fixed by the two vacuum equations. Let \(\cH_a\) collect the holographic data. The calculated path is
\begin{equation}
 \Theta_{\rm str}\xrightarrow{\ {\rm KK}\ }\cI_A
 \xrightarrow{\ {\rm HR}\ }
 \cH_a=\{a_E,a_C,\kappa,C_g,C_\phi,D_g,D_\phi,\Delta_T,\Delta_S\}.
 \label{eq:intro-inference-map}
\end{equation}
The closed formulas below determine the exact holographic Jacobian and its reduction pullback,
\begin{equation}
 J^{\rm HR}_{aA}=\frac{\partial\cH_a}{\partial\cI_A},
 \qquad
 J^{\rm red}_{aI}=J^{\rm HR}_{aA}\frac{\partial\cI_A}{\partial\Theta_{\rm str}^I}.
 \label{eq:intro-holographic-jacobian}
\end{equation}
For each specified compactification map \(\Theta_{\rm str}\mapsto\cI\), the row space of \(J^{\rm red}\) identifies the compactification directions resolved by anomaly data, normalizable mixing, response determinants, and local source currents. A gravitational observable map \(\cP_\mu(\cI)\) contributes the block \(J^{\rm grav}_{\mu A}=\partial\cP_\mu/\partial\cI_A\) to the same joint inference matrix. Equations~\eqref{eq:intro-inference-map} and \eqref{eq:intro-holographic-jacobian} furnish the exact holographic block and the coefficient coordinates shared with black-object and propagation analyses~\cite{LIGO2017,EzquiagaZumalacarregui2017}.

On the regular scalar-normalization quotient, we obtain a global rational inverse for the invariant quartet \((C_g,sC_\phi,s^2D_\phi,a_E+a_C)\) with Jacobian determinant \(128/\ell\). The response block alone leaves one coupling direction unresolved, and the curvature-anomaly sum removes this kernel exactly. After the triangular fixed-\(n\) Kaluza--Klein pullback, \(\rank J^{\rm red}=4\) and \(\ker J^{\rm red}=\{0\}\), with \(\widetilde\alpha_{\rm GB}=-2(a_E+a_C)/\ell\).

The paper is organized as follows. Section~\ref{sec:sdlh-origin} develops the SDLH coupling structure and the linear-dilaton AlAdS branch from the string/KK construction. Section~\ref{sec:radial-variation} derives the radial Gauss--Codazzi decomposition and the complete Dirichlet variational density. Section~\ref{sec:fg-recursion} solves the boundary-covariant weight-two and weight-four Fefferman--Graham recursion. Section~\ref{sec:counterterms-anomaly} constructs the counterterms, exact anomaly, and logarithmic obstruction. Section~\ref{sec:onepoints} extracts the finite stress tensor and scalar one-point function. Section~\ref{sec:ward} derives the momentum and trace Ward identities from independent radial constraints. Section~\ref{sec:implications} develops the response-channel and string-to-observable maps, and section~\ref{sec:conclusions} collects the resulting physical picture.

\section{The string-derived Lovelock--Horndeski theory and the linear-dilaton branch}
\label{sec:sdlh-origin}

\subsection{Coefficient frames and dimensional reduction}

The metric--dilaton sector of the heterotic effective action supplies the higher-dimensional starting point. At first order in \(\alpha'\), a convenient coefficient-frame representative is
\begin{equation}
\begin{aligned}
\widehat S=\frac{1}{16\pi G_D}\int \dd^Dx\sqrt{-\widehat G}\,
e^{-2\widehat\Phi}\Big\{&\widehat R+4(\widehat\nabla\widehat\Phi)^2
+\alpha'\Big( \widehat R_{ABCD}\widehat R^{ABCD}
+b_1\widehat R_{AB}\widehat R^{AB}+b_2\widehat R^2\\
&+b_3\widehat R^{AB}\widehat\Phi_A\widehat\Phi_B
+b_4\widehat R(\widehat\nabla\widehat\Phi)^2
+b_5\widehat R\widehat\Box\widehat\Phi
+b_6(\widehat\Box\widehat\Phi)^2\\
&+b_7(\widehat\nabla\widehat\Phi)^2\widehat\Box\widehat\Phi
+b_8(\widehat\nabla\widehat\Phi)^4\Big) \bigg\}.
\end{aligned}
\label{eq:heterotic-frame}
\end{equation}
Here \(A,B=0,\ldots,D-1\), \(\widehat\Phi_A=\widehat\nabla_A\widehat\Phi\), and the coefficients label local field variables inside the order-\(\alpha'\) equivalence class~\cite{MetsaevTseytlin1987,GrossSloan1987,BergshoeffdeRoo1989,ChemissanydeRooPanda2007}. Local metric and dilaton redefinitions move the \(b_I\) while preserving on-shell amplitudes. The coefficient-frame operators and reduction polynomials are given in~\cite{WuStone2026}. We select the Gauss--Bonnet--Horndeski representative \((b_1,\ldots,b_8)=\left(-4,1,a_1,a_2,0,0,a_3,a_4\right)\), where \(a_2=-a_1/2\). Its overall parent coupling is \(\widetilde\alpha_{\rm GB}\equiv\alpha'\), and its independent parent parameters \((\widetilde\alpha_{\rm GB},a_1,a_3,a_4)\) map under KK reduction to the five-dimensional couplings \((\alpha_1,\alpha_2,\alpha_3,\alpha_4)\) below. The same map pulls the holographic observables back to the higher-dimensional coefficient frame at the matching scale. A frame change simultaneously transports sources and finite local contact terms, so the generating functional is the natural comparison object.

For a maximally symmetric internal space \(\mathcal K_n\), take
\begin{equation}
 \dd\widehat s_D^2=e^{2\alpha\phi(x)}\dd s_5^2
 +e^{2\beta\phi(x)}\dd\Omega_n^2,
 \qquad \widehat\Phi=\kappa_\Phi\phi,
 \qquad D=5+n .
 \label{eq:kk-ansatz}
\end{equation}
The Einstein-frame relation is \(3\alpha+n\beta-2\kappa_\Phi=0\). Substitution of Eq.~\eqref{eq:kk-ansatz}, followed by covariant integrations by parts, populates precisely the five operators in Eq.~\eqref{eq:intro-action}: the leading scalar kinetic term, the Gauss--Bonnet density, the Einstein-tensor kinetic interaction, the cubic Galileon, and the quartic scalar kinetic term~\cite{DuffNilssonPope1986, WuStone2026,VanAcoleyenVanDoorsselaere2011}.

For the reduction trajectory of~\cite{WuStone2026}, the choice
\begin{equation}
 \alpha=0,\qquad \lambda_{\mathcal K}=0,\qquad n\beta=2,
 \qquad \widetilde\alpha_{\rm GB}\ne0,
 \label{eq:kk-trajectory}
\end{equation}
sets one warp exponent and the internal curvature. It retains the full curvature-squared interaction and gives the explicit five-dimensional map
\begin{align}
 \alpha_1={}&\widetilde\alpha_{\rm GB},
 \notag\\
 \alpha_2={}&\widetilde\alpha_{\rm GB}
 \left[a_1-4n(n-1)\beta^2\right],
 \notag\\
 \alpha_3={}&\widetilde\alpha_{\rm GB}
 \left[\frac{3}{2}a_1n\beta-2n(n-1)(n-2)\beta^3+a_3\right],
 \notag\\
 \alpha_4={}&\widetilde\alpha_{\rm GB}
 \left[n(n-1)^2(n-2)\beta^4-\frac{1}{2}a_1n(n-3)\beta^3+a_4\right].
 \label{eq:explicit-kk-coupling-map}
\end{align}
Here \(a_1,a_3,a_4\) are the coefficient-frame parameters inherited from the higher-dimensional derivative operators. We fix \(\alpha_0=-4/5\), while its symbolic dependence makes the scalar-normalization orbit and the final quotient manifest. Equation~\eqref{eq:explicit-kk-coupling-map} displays \(\alpha_1=\widetilde\alpha_{\rm GB}\) directly and fixes the coefficient correspondence used throughout the holographic calculation.

A real nonzero scalar rescaling \(\phi=\lambda\varphi\) traces an exact trajectory through the five-dimensional coefficient space:
\begin{equation}
 (\alpha_0,\alpha_1,\alpha_2,\alpha_3,\alpha_4;s)
 \longmapsto
 (\lambda^2\alpha_0,\alpha_1,\lambda^2\alpha_2,
 \lambda^3\alpha_3,\lambda^4\alpha_4;s/\lambda).
 \label{eq:scalar-normalization-trajectory}
\end{equation}
The normalization-invariant coordinates are
\begin{equation}
 \mathsf c_0=\alpha_0s^2,\qquad
 \mathsf c_1=\frac{\alpha_1}{\ell^2},\qquad
 \mathsf c_2=\frac{\alpha_2s^2}{\ell^2},\qquad
 \mathsf c_3=\frac{\alpha_3s^3}{\ell^2},\qquad
 \mathsf c_4=\frac{\alpha_4s^4}{\ell^2}.
 \label{eq:scalar-normalization-invariants}
\end{equation}
They coordinatize the scalar-normalization quotient of the regular response manifold and the closed string-coefficient inverse in Sec.~\ref{subsec:closed-string-inverse}.

The shift symmetry is manifest at the five-dimensional level. Set \(\phi_M=\nabla_M\phi\) and \(\phi_{MN}=\nabla_M\nabla_N\phi\). The scalar equation is the conservation law
\begin{equation}
 \cE_\phi=\nabla_MJ^M=0,
 \qquad
 J^M=\alpha_0\phi^M+\alpha_2\cG^{MN}\phi_N
 +\alpha_3\!\Big(\phi^M\Box\phi-\phi_N\phi^{MN}\Big)
 +2\alpha_4X\phi^M .
 \label{eq:shift-current}
\end{equation}
Using \(\nabla_M\cG^{MN}=0\) and commuting derivatives gives the explicit second-order equation
\begin{equation}
\begin{aligned}
\cE_\phi={}&\alpha_0\Box\phi+\alpha_2\cG^{MN}\phi_{MN}
+\alpha_3 \!\Big( (\Box\phi)^2-\phi_{MN}\phi^{MN}-R_{MN}\phi^M\phi^N\Big)\\
&+2\alpha_4\!\Big( X\Box\phi+2\phi^M\phi^N\phi_{MN}\Big) =0.
\end{aligned}
\label{eq:scalar-eom}
\end{equation}
Our scalar equation uses the current-divergence normalization
\begin{equation}
 \frac{16\pi G_5}{\sqrt{-G}}\frac{\delta S_{\rm bulk}}{\delta\phi}
 \equiv-2\cE_\phi .
 \label{eq:scalar-eom-normalization}
\end{equation}
The metric equation can be organized sectorwise as
\begin{equation}
 \cE_{MN}=\cG_{MN}
 +\alpha_0\!\left(\phi_M\phi_N-\frac12G_{MN}X\right)
 +\alpha_1\cH^{\rm GB}_{MN}-\alpha_2\Theta_{MN}
 +\alpha_3\cC_{MN}+\alpha_4\cQ_{MN}=0,
 \label{eq:metric-eom-compact}
\end{equation}
with
\begin{align}
\cH^{\rm GB}_{MN}={}&2RR_{MN}-4R_{MP}R_N{}^P-4R^{PQ}R_{MPNQ}\notag\\
&+2R_M{}^{PQR}R_{NPQR}-\frac12G_{MN}\cL_{\rm GB},
\label{eq:hgb}\\
\Theta_{MN}={}&\frac12R\phi_M\phi_N-R_{MP}\phi^P\phi_N-R_{NP}\phi^P\phi_M\notag\\
&-R_{MPNQ}\phi^P\phi^Q+\phi_{MN}\Box\phi-\phi_{MP}\phi_N{}^P
+\frac12\cG_{MN}X\notag\\
&-G_{MN}\!\left[\frac12(\Box\phi)^2-\frac12\phi_{PQ}\phi^{PQ} -R_{PQ}\phi^P\phi^Q\right],
\label{eq:theta-mn}\\
\cC_{MN}={}&\phi_M\phi_N\Box\phi
-\left(\phi_M\phi_{NP}+\phi_N\phi_{MP}\right)\phi^P
 +G_{MN}\phi^P\phi^Q\phi_{PQ},\nonumber\\
\cQ_{MN}={}&2X\phi_M\phi_N-\frac12G_{MN}X^2.
\label{eq:cq-mn}
\end{align}
Thus the string reduction preserves the five-dimensional Lovelock--Horndeski second-order structure in Eqs.~\eqref{eq:scalar-eom} and \eqref{eq:metric-eom-compact}.

\subsection{Vacuum polynomials and exact response domain}

The linear-dilaton branch is most transparent in FG coordinates,
\begin{equation}
 \dd s^2=\frac{\ell^2}{4\rho^2}\dd\rho^2+\rho^{-1}\eta_{ij}\dd x^i\dd x^j,
 \qquad \phi=s\log\rho+\phi_c .
\label{eq:linear-dilaton-background}
\end{equation}
On this background the elementary invariants are
\begin{align}
 R_{MN}&=-\frac{4}{\ell^2}G_{MN},
 &R&=-\frac{20}{\ell^2},
 &\cG_{MN}&=\frac{6}{\ell^2}G_{MN},
 \notag\\
 \cL_{\rm GB}&=\frac{120}{\ell^4},
 &X&=\frac{4s^2}{\ell^2},
 &\Box\phi&=-\frac{8s}{\ell^2},
 \label{eq:background-invariants}
\end{align}
with \(\nabla_\rho\nabla_\rho\phi=0\) and
\(\nabla_i\nabla_j\phi=-2sG_{ij}/\ell^2\).
Substitution into Eqs.~\eqref{eq:scalar-eom} and \eqref{eq:metric-eom-compact} gives two independent vacuum polynomials,
\begin{align}
V_1&=24\alpha_1-24\alpha_2s^2+16\alpha_4s^4
+\ell^2\left(4\alpha_0s^2-12\right)=0,
\label{eq:vacuum-v1}\\
V_2&=\frac12\alpha_0\ell^2+3\alpha_2-4\alpha_3s+4\alpha_4s^2=0.
\label{eq:vacuum-v2}
\end{align}
The AdS radius is therefore generated by the derivative sectors and the radial scalar slope. Solving the two branch equations for \(\alpha_1\) and \(\alpha_4\) gives
\begin{equation}
 \alpha_4=\frac{-\alpha_0\ell^2-6\alpha_2+8\alpha_3s}{8s^2},
 \qquad
 \alpha_1=\frac{\ell^2}{2}+\frac{3}{2}\alpha_2s^2
 -\frac{1}{12}\alpha_0\ell^2s^2-\frac{2}{3}\alpha_3s^3 .
 \label{eq:branch-elimination}
\end{equation}
The pair \((\ell,s)\) now acts as branch data; \(\alpha_0,\alpha_2,\alpha_3\) provide a convenient exact-coupling chart. The regular branch is the algebraic response manifold
\begin{equation}
 \mathcal M_{\rm reg}=
 \left\{(\alpha_0,\ldots,\alpha_4;\ell,s):
 V_1=V_2=0,\ s\Delta_T\Delta_S\ne0\right\}.
 \label{eq:regular-branch-manifold}
\end{equation}
Solving \((V_1,V_2)\) for \((\alpha_1,\alpha_4)\) gives the five-dimensional chart
\((\alpha_0,\alpha_2,\alpha_3;\ell,s)\). Quotienting the scalar-normalization trajectory \eqref{eq:scalar-normalization-trajectory} gives a four-dimensional invariant response manifold coordinatized by four independent combinations drawn from Eq.~\eqref{eq:scalar-normalization-invariants} together with the branch equations.

Two combinations control the radial response:
\begin{align}
 \Delta_T&=3\ell^2+12\alpha_2s^2-\alpha_0\ell^2s^2-8\alpha_3s^3,
 \label{eq:delta-t}\\
 \Delta_S&=18\alpha_2+3\alpha_0\ell^2-12\alpha_3s
 +6\alpha_0\alpha_2s^2-\alpha_0^2\ell^2s^2-4\alpha_0\alpha_3s^3.
 \label{eq:delta-s}
\end{align}
We use the analytic response chart
\begin{equation}
 s\Delta_T\Delta_S\ne0.
 \label{eq:regular-response-domain}
\end{equation}
The hypersurfaces \(\Delta_T=0\) and \(\Delta_S=0\) are resonant tensor and scalar/trace channels where the indicial kernel enlarges. The recursion therefore identifies them as the coupling-space loci where the logarithmic response modes reorganize.

Spatially varying scalar sources preserve the leading AlAdS geometry. The asymptotic power count reads
\begin{equation}
 G^{\rho\rho}(\partial_\rho\phi)^2=\frac{4s^2}{\ell^2}+O(\rho),
 \qquad
 G^{ij}\partial_i\phi\partial_j\phi
 =\rho\,g_{(0)}^{ij}\partial_i\phi_{(0)}\partial_j\phi_{(0)}+O(\rho^2).
 \label{eq:source-power-counting}
\end{equation}
Thus \(s\) fixes the leading branch and \(\phi_{(0)}(x)\) enters the backreaction at derivative weight two. This separation supplies the asymptotic hierarchy used in the radial variational problem.

\section{Radial variation and Dirichlet completion}
\label{sec:radial-variation}

\subsection{FG slicing as radial ADM geometry}

FG gauge is a radial ADM slicing with
\begin{equation}
 N=\frac{\ell}{2\rho},\qquad N^i=0,\qquad
 \gamma_{ij}=\rho^{-1}g_{ij},\qquad
 \dd s^2=N^2\dd\rho^2+\gamma_{ij}\dd x^i\dd x^j .
 \label{eq:radial-adm}
\end{equation}
Choose the outward normal at the ultraviolet cutoff as \(n_{\rm out}=-2\rho\partial_\rho/\ell\). The intrinsic derivative \(D_i\), extrinsic curvature, normal scalar velocity, and tangential scalar gradient are
\begin{equation}
 K_{ij}=\frac12\cL_n\gamma_{ij},\qquad
 K_{ij}^{\rm out}=\frac{1}{\ell}\left(\rho^{-1}g_{ij}-g'_{ij}\right),
 \qquad v=n^M\nabla_M\phi,\qquad X_i=D_i\phi .
 \label{eq:radial-data}
\end{equation}
The Gauss--Codazzi relations decompose bulk curvature into intrinsic and radial data,
\begin{align}
{}^{(5)}R_{ijkl}&=\widehat R_{ijkl}-K_{ik}K_{jl}+K_{il}K_{jk},
\label{eq:gauss}\\
{}^{(5)}R_{\perp ijk}&=D_jK_{ik}-D_kK_{ij},
\qquad
\cG_{\perp i}=D_jK^j{}_i-D_iK,
\label{eq:codazzi}\\
{}^{(5)}R&=\widehat R+K^2-K_{ij}K^{ij}
-2\nabla_M\!\left(Kn^M-a^M\right),
\qquad a_i=0 .
\label{eq:scalar-gauss-codazzi}
\end{align}
For a flat pure-AdS slice, \(K_{ij}^{\rm out}=\gamma_{ij}/\ell\) and \(\widehat R_{ijkl}=0\), so Eq.~\eqref{eq:gauss} gives the projected curvature \(-\ell^{-2}(\gamma_{ik}\gamma_{jl}-\gamma_{il}\gamma_{jk})\). Hats denote tensors of \(\gamma_{ij}\). The scalar derivatives decompose through \(X=X_kX^k+v^2\) and
\begin{equation}
 \Box_5\phi=D^2\phi+Kv+\cL_nv-a^iX_i .
 \label{eq:box-decomposition}
\end{equation}
We use Eqs.~\eqref{eq:gauss}--\eqref{eq:box-decomposition} to expose the normal derivatives entering the bulk variation and to identify the boundary density required by Dirichlet data.

\subsection{Sectorwise derivation of the surface density}

We take each regulator hypersurface to be smooth and closed. Define the Dirichlet surface functional and completed action by
\begin{equation}
 S_{\partial,D}=\frac{1}{16\pi G_5}\int_{\partial\mathcal M}\dd^4x
 \sqrt{-\gamma}\,\cB_D,
 \qquad S_{\rm comp}=S_{\rm bulk}+S_{\partial,D} .
 \label{eq:completed-action}
\end{equation}
The Einstein--Hilbert variation generates the familiar normal derivative of \(\delta\gamma_{ij}\), and \(2K\) converts it into the Brown--York canonical momentum~\cite{GibbonsHawking1977,BrownYork1993}. The Gauss--Bonnet variation is the second Lovelock boundary polynomial: its normal-derivative terms assemble into the Myers density \(4\alpha_1(J-2\widehat\cG_{ij}K^{ij})\)~\cite{Myers1987,Davis2003,GravanisWillison2003}. For the Einstein-tensor kinetic sector, the Horndeski representative \(G_5=-\alpha_2\phi\) is related to the John interaction by
\begin{equation}
 -\alpha_2\phi\,\cG^{MN}\nabla_M\nabla_N\phi
 =\alpha_2\cG^{MN}\nabla_M\phi\nabla_N\phi
 -\alpha_2\nabla_M\!\left(\phi\cG^{MN}\nabla_N\phi\right),
 \label{eq:john-representative-identity}
\end{equation}
where \(\nabla_M\cG^{MN}=0\). Combining its boundary primitive with the displayed divergence and one tangential integration by parts yields \(\alpha_2(K^{ij}X_iX_j-KX_kX^k)\). The cubic interaction follows from \(\cL_3=-G_3\Box\phi\), \(G_3=2\alpha_3X_H\), \(X_H=-X/2\), \(Y=-X_kX^k/2\), and
\begin{equation}
 F_3=\int_0^vG_3\!\left(\phi,Y-\frac{x^2}{2}\right)\dd x
 =-\alpha_3vX_kX^k-\frac{\alpha_3}{3}v^3,
 \label{eq:galileon-boundary-primitive}
\end{equation}
which is the scalar boundary primitive in the general Horndeski construction~\cite{PadillaSivanesan2012}. The first-derivative \(\alpha_0X\) and \(\alpha_4X^2\) sectors already possess Dirichlet variations.

We combine the four contributions into
\begin{equation}
 \cB_D=2K+4\alpha_1\left(J-2\widehat\cG_{ij}K^{ij}\right)
 +\alpha_2\left(K^{ij}X_iX_j-KX_kX^k\right)
 -\alpha_3v\left(X_kX^k+\frac{v^2}{3}\right).
 \label{eq:dirichlet-density}
\end{equation}
The cubic extrinsic-curvature tensor is
\begin{equation}
 J_{ij}=\frac{1}{3}\left(2KK_{ik}K^k{}_j+K_{kl}K^{kl}K_{ij}
 -2K_{ik}K^{kl}K_{lj}-K^2K_{ij}\right),
 \qquad J=J^i{}_i .
 \label{eq:myers-j}
\end{equation}
We obtain the defining variation in canonical form:
\begin{equation}
 \left.\delta S_{\rm comp}\right|_{\partial\mathcal M}
 =\frac{1}{16\pi G_5}\int\dd^4x\sqrt{-\gamma}
 \left(\Pi^{ij}\delta\gamma_{ij}+\Pi_\phi\delta\phi+D_iY^i\right).
 \label{eq:dirichlet-variation}
\end{equation}
The coefficients of \(\cL_n\delta\gamma_{ij}\) and \(\delta v\) cancel sector by sector, leaving \((\gamma_{ij},\phi)\) as the boundary configuration variables and \((\Pi^{ij},\Pi_\phi)\) as their radial canonical responses.

We use the mixed-index canonical-current convention
\begin{equation}
 \cJ_D{}^i{}_j\equiv\Pi^{ik}\gamma_{kj},
 \qquad \cJ_D^\phi\equiv\Pi_\phi .
 \label{eq:canonical-current-definition}
\end{equation}

\subsection{Completed cutoff currents}

Define the double-dual intrinsic curvature
\begin{equation}
 \widehat P_{ikjl}=\widehat R_{ikjl}+\widehat R_{kj}\gamma_{il}
 -\widehat R_{kl}\gamma_{ij}-\widehat R_{ij}\gamma_{kl}
 +\widehat R_{il}\gamma_{kj}
 +\frac{1}{2}\widehat R(\gamma_{ij}\gamma_{kl}-\gamma_{il}\gamma_{kj}).
 \label{eq:double-dual}
\end{equation}
Direct variation of Eq.~\eqref{eq:completed-action} yields the mixed-index metric current
\begin{equation}
\begin{aligned}
\cJ_D{}^i{}_j={}&K\delta^i_j-K^i{}_j-6\alpha_1J^i{}_j
+2\alpha_1J\delta^i_j+4\alpha_1\widehat P^i{}_{kjl}K^{kl}\\
&+\alpha_2\bigg[\frac{v^2}{2}(K\delta^i_j-K^i{}_j)
+\frac{X_kX^k}{2}(K^i{}_j-K\delta^i_j)
+K_{kl}X^kX^l\delta^i_j+KX^iX_j\\
&\hspace{3.4em}-X^iK_{jk}X^k-X_jK^i{}_kX^k
+v(H\delta^i_j-H^i{}_j)\bigg]
+\alpha_3\left(vX^iX_j+\frac{v^3}{3}\delta^i_j\right),
\end{aligned}
\label{eq:dirichlet-metric-current}
\end{equation}
where \(H_{ij}=D_iD_j\phi\) and \(H=D^2\phi\). The scalar current is
\begin{equation}
\begin{aligned}
\cJ_D^\phi={}&2\alpha_0v+4\alpha_4(X_kX^k+v^2)v
+2\alpha_3\left(2vH+Kv^2+K_{ij}X^iX^j\right)\\
&+\alpha_2v\left(K^2-K_{ij}K^{ij}-\widehat R\right)
-2\alpha_2\left(K^{ij}H_{ij}-KH\right).
\end{aligned}
\label{eq:dirichlet-scalar-current}
\end{equation}
These are the finite-cutoff currents whose renormalized limits define \(\langle T_{ij}\rangle\) and \(\langle\cO_\phi\rangle\). Their common variational origin in Eq.~\eqref{eq:completed-action} enforces reciprocal metric--scalar mixing and carries the radial constraints into the boundary Noether identities.

The bulk equations split into
\begin{equation}
 \cE_{\rho\rho}=0,\qquad \cE_{\rho i}=0,\qquad
 \cE_{ij}=0,\qquad \cE_\phi=0.
 \label{eq:eom-split}
\end{equation}
The first component controls the radial Hamiltonian relation, the four \(\cE_{\rho i}\) components control momentum exchange, the tangential tensor equation determines metric response, and the scalar equation determines scalar response. This decomposition supplies the FG recursion with a consistent radial variational system.

\section{The Fefferman--Graham recursion}
\label{sec:fg-recursion}

The Fefferman--Graham expansion resolves the covariant equations into a finite sequence of tensor systems~\cite{HenningsonSkenderis1998,deHaroSkenderisSolodukhin2001,Skenderis2002}. The complete source and response ansatz is Eq.~\eqref{eq:intro-fg}. The pair \((g_{(0)ij},\phi_{(0)})\) supplies arbitrary sources; \((g_{(2)ij},\phi_{(2)})\) is locally determined at weight two; \((g_{(4)ij}^{\rm tot},\phi_{(4)}^{\rm tot})\) contains the normalizable response; and \((h_{(4)ij},\psi_{(4)})\) carries the logarithmic obstruction.

\subsection{Radial identities and Laurent--log algebra}
\label{subsec:series-algebra}

Introduce \(A^i{}_j=(g^{-1}g')^i{}_j\), where a prime denotes \(\partial_\rho\). Direct evaluation of the Christoffel symbols of Eq.~\eqref{eq:intro-fg} gives
\begin{align}
R_{\rho i}&=\frac12\left(D^jg'_{ij}-D_i\tr A\right),
\label{eq:rho-i-identity}\\
R_{\rho\rho}&=-\rho^{-2}-\frac12\tr(g^{-1}g'')
+\frac14\tr(A^2),
\label{eq:rho-rho-identity}\\
\Box_5\phi&=\frac{4}{\ell^2}\left[
\rho^2\phi''-\rho\phi'+\frac{\rho^2}{2}\tr A\,\phi'\right]
+\rho\Box_g\phi,
\label{eq:box-fg}\\
X&=\frac{4\rho^2}{\ell^2}(\phi')^2
+\rho g^{ij}D_i\phi D_j\phi,
\qquad
\sqrt{-G}=\frac{\ell}{2}\rho^{-3}\sqrt{-g}.
\label{eq:x-det-fg}
\end{align}
These identities retain the full boundary covariant derivative and therefore apply to arbitrary \(g_{(0)ij}(x)\) and \(\phi_{(0)}(x)\).

For a Laurent--log series
\begin{equation}
 F(\rho)=\sum_{p,q}F_{p,q}\rho^p(\log\rho)^q,
 \label{eq:formal-series}
\end{equation}
radial differentiation and multiplication obey
\begin{align}
\partial_\rho\!\left[\rho^p(\log\rho)^q\right]
&=p\rho^{p-1}(\log\rho)^q
+q\rho^{p-1}(\log\rho)^{q-1},
\label{eq:log-derivative}\\
(FG)_{p,q}
&=\sum_{\substack{r+s=p\\m+n=q}}F_{r,m}G_{s,n}.
\label{eq:series-convolution}
\end{align}
Writing \(B=g^{-1}\), the inverse metric is generated recursively by
\begin{equation}
 B_{0,0}=g_{(0)}^{-1},\qquad
 B_{p,q}=-g_{(0)}^{-1}
 \sum_{(r,m)\ne(0,0)}g_{r,m}B_{p-r,q-m},
 \label{eq:inverse-series}
\end{equation}
and its determinant follows from
\begin{equation}
 \partial_\rho\log\sqrt{-g}=\frac{1}{2}\tr(g^{-1}g').
 \label{eq:det-series}
\end{equation}
Radial normalization aligns the scalar, Hamiltonian, and tangential equations at a common Laurent weight:
\begin{equation}
 \overline\cE_{\rho\rho}=\rho^2\cE_{\rho\rho},\qquad
 \overline\cE^i{}_j=\rho g^{ik}\cE_{kj},\qquad
 \overline\cE_{\rho i}=\cE_{\rho i},\qquad
 \overline\cE_\phi=\cE_\phi .
 \label{eq:normalized-eom-blocks}
\end{equation}
Each normalized field equation is then projected through
\begin{equation}
 \overline\cE_A^{[p,q]}\equiv
 \operatorname{Coeff}_{\rho^p(\log\rho)^q}\overline\cE_A,
 \qquad
 w=2p\quad(A=\rho\rho,i{\,}^{j},\phi),
 \qquad w=2p+3\quad(A=\rho i) .
 \label{eq:coefficient-extractor}
\end{equation}
We project the leading equations to establish the branch, while the weight-two rows determine
local backreaction. At weight four the \([2,1]\) row places the logarithmic
coefficient in the resonant kernel, while the solvability projection of the
\([2,0]\) row fixes its amplitude. The complementary finite rows determine
the local particular solution together with the trace, divergence, and scalar
constraints on the normalizable state response. The result is a
finite-dimensional covariant system at every derivative weight.

\subsection{Arbitrary sources and the exact weight-two solution}
\label{subsec:weight-two}

All tensors in this subsection are constructed from \(g_{(0)ij}\). The intrinsic scalar-source data are
\begin{equation}
 X_i=D_i\phi_{(0)},\qquad H_{ij}=D_iD_j\phi_{(0)},
 \qquad H=D^2\phi_{(0)} .
 \label{eq:boundary-jets}
\end{equation}
Parity and shift symmetry give the complete weight-two basis
\begin{align}
g_{(2)ij}={}&c_{\rm Ric}R_{ij}+c_RRg_{(0)ij}
+c_{XX}X_iX_j+c_{X^2}X_kX^kg_{(0)ij}
+c_HH_{ij}+c_{\Box}Hg_{(0)ij},
\label{eq:g2-ansatz}\\
\phi_{(2)}={}&b_RR+b_{X^2}X_kX^k+b_{\Box}H.
\label{eq:phi2-ansatz}
\end{align}
Projecting \(\overline\cE^i{}_j{}^{[1,0]}\), \(\overline\cE_{\rho\rho}^{[1,0]}\), and \(\overline\cE_\phi^{[1,0]}\) onto independent curvature and scalar jets yields
\begin{equation}
 M_2\bm x_2=\bm r_2,\qquad
 \bm x_2=(c_{\rm Ric},c_R,c_{XX},c_{X^2},c_H,c_{\Box},b_R,b_{X^2},b_{\Box})^{\mathsf T}.
 \label{eq:weight-two-system}
\end{equation}
The retained row order is explicit:
\begin{equation}
\begin{aligned}
 \bm\Pi_2=\big(&
 [\overline\cE^i{}_j]_{R^i{}_j},
 [\overline\cE^i{}_j]_{\delta^i_jR},
 [\overline\cE^i{}_j]_{X^iX_j},\\
&[\overline\cE^i{}_j]_{\delta^i_jX_kX^k},
 [\overline\cE^i{}_j]_{H^i{}_j},
 [\overline\cE^i{}_j]_{\delta^i_jH};\\
&[\overline\cE_{\rho\rho}]_{R,X_kX^k,H};
 [\overline\cE_\phi]_{R,X_kX^k,H}
 \big)^{[1,0]} .
\end{aligned}
 \label{eq:weight-two-projection-order}
\end{equation}
Thus \(M_2\) is a \(12\times9\) exact rational matrix in the declared ordering. A nonvanishing regular-branch rank minor and the augmented rank give
\begin{equation}
 \exists\ I_9\subset\{1,\ldots,12\}:\quad
 \det M_2[I_9,:]\ne0,
 \qquad
 \operatorname{rank}M_2=\operatorname{rank}(M_2\mid\bm r_2)=9.
 \label{eq:weight-two-rank-certificate}
\end{equation}
On the response domain \eqref{eq:regular-response-domain},
\begin{equation}
 \operatorname{rank}M_2=\operatorname{rank}(M_2\mid\bm r_2)=9,
 \qquad \dim\ker M_2=0,
 \qquad M_2\bm x_2-\bm r_2=0.
 \label{eq:weight-two-rank}
\end{equation}
Twelve retained projections therefore determine the nine coefficients and simultaneously furnish three compatibility identities.

Introduce the branch polynomials
\begin{align}
U={}&3\ell^2+24\alpha_2s^2-\alpha_0\ell^2s^2-8\alpha_3s^3,
\label{eq:u-polynomial}\\
V={}&6\ell^2+18\alpha_2s^2-\alpha_0\ell^2s^2-8\alpha_3s^3,
\label{eq:v-polynomial}\\
N={}&-3\alpha_3\ell^2+18\alpha_2^2s-24\alpha_2\alpha_3s^2
+\alpha_0\alpha_3\ell^2s^2+8\alpha_3^2s^3,
\label{eq:n-polynomial}\\
C={}&27\alpha_2+3\alpha_0\ell^2-18\alpha_3s
+3\alpha_0\alpha_2s^2-\alpha_0^2\ell^2s^2
-2\alpha_0\alpha_3s^3,
\label{eq:c-polynomial}
\end{align}
and the two numerators
\begin{align}
N_{X^2}={}&-54\alpha_2^2\ell^2-9\alpha_0\alpha_2\ell^4
+54\alpha_2\alpha_3\ell^2s-108\alpha_2^3s^2
-18\alpha_0\alpha_2^2\ell^2s^2-12\alpha_3^2\ell^2s^2\notag\\
&+3\alpha_0^2\alpha_2\ell^4s^2+216\alpha_2^2\alpha_3s^3
+6\alpha_0\alpha_2\alpha_3\ell^2s^3-144\alpha_2\alpha_3^2s^4
+4\alpha_0\alpha_3^2\ell^2s^4+32\alpha_3^3s^5,
\label{eq:nq-polynomial}\\
N_B={}&-18\alpha_2\ell^2-3\alpha_0\ell^4+18\alpha_3\ell^2s
-36\alpha_2^2s^2-6\alpha_0\alpha_2\ell^2s^2+\alpha_0^2\ell^4s^2\notag\\
&+48\alpha_2\alpha_3s^3+2\alpha_0\alpha_3\ell^2s^3-16\alpha_3^2s^4.
\label{eq:nb-polynomial}
\end{align}
The exact solution is
\begin{align}
c_{\rm Ric}&=-\frac{\ell^2U}{2\Delta_T},
&c_R&=\frac{\ell^2UC-(3\alpha_2-2\alpha_3s)\Delta_T^2
}{12\Delta_T\Delta_S},
\notag\\
c_{XX}&=-\frac{3\alpha_2\ell^2}{\Delta_T},
&c_{X^2}&=-\frac{N_{X^2}}{2\Delta_T\Delta_S},
\notag\\
c_H&=\frac{6\alpha_2\ell^2s}{\Delta_T},
&c_{\Box}&=\frac{2s^2(3\alpha_2-2\alpha_3s)N}{\Delta_T\Delta_S},
\notag\\
b_R&=\frac{s^2N}{6\Delta_S},
&b_{X^2}&=\frac{N_B}{8s\Delta_S},
&b_{\Box}&=-\frac{sN}{2\Delta_S}.
\label{eq:weight-two-solution}
\end{align}
Our exact solution in Eq.~\eqref{eq:weight-two-solution} converts arbitrary boundary curvature, scalar gradients, and scalar Hessians into the complete local backreaction. The simultaneous appearance of \(\Delta_T\) and \(\Delta_S\) anticipates their role in logarithmic response and finite metric--scalar mixing.

\subsection{Weight-four local, state, and logarithmic sectors}
\label{subsec:weight-four}

The total weight-four coefficients admit the split
\begin{equation}
 g_{(4)ij}^{\rm tot}=g_{(4)ij}^{\rm local}+g_{(4)ij}^{\rm state},
 \qquad
 \phi_{(4)}^{\rm tot}=\phi_{(4)}^{\rm local}+\phi_{(4)}^{\rm state}.
 \label{eq:weight-four-split}
\end{equation}
The source-local pieces in Eq.~\eqref{eq:weight-four-split} are fixed covariant functionals of \(g_{(0)}\) and \(\phi_{(0)}\). The state pieces lie in the homogeneous kernel of the weight-four radial operator and are selected by the regular interior bulk saddle. Their trace and divergence obey the radial constraints, while their unconstrained normalizable components carry the state dependence of the one-point functions. To define the radial indicial blocks, suppress boundary derivatives and perturb the weight-four response by
\begin{equation}
 \delta g_{ij}=\rho^\lambda
 \left(t_{ij}^{\rm TT}+\frac{1}{4}g_{(0)ij}\tau\right),
 \qquad
 \delta\phi=\rho^\lambda\varphi,
 \qquad
 \bm y_{\rm S}=(\tau,\varphi)^{\mathsf T}.
 \label{eq:indicial-variables}
\end{equation}
The transverse-traceless projection of \(\overline\cE^i{}_j\) defines the scalar polynomial \(P_{\rm TT}(\lambda)\). The trace of \(\overline\cE^i{}_j\) together with \(\overline\cE_\phi\) defines the \(2\times2\) block \(P_{\rm scalar/trace}(\lambda)\) acting on \(\bm y_{\rm S}\). After imposing \(V_1=V_2=0\), their invariant factors are
\begin{equation}
 P_{\rm TT}(\lambda)\propto\lambda(\lambda-2)\Delta_T,
 \qquad
 \det P_{\rm scalar/trace}(\lambda)
 =-\frac{8}{\ell^4}(\lambda-2)\lambda^2\Delta_S.
 \label{eq:indicial-factors}
\end{equation}
For a radial operator \(\mathsf D\),
\begin{equation}
 \mathsf D[\rho^\lambda\bm y]=\rho^\lambda P(\lambda)\bm y,
 \qquad
 \mathsf D[\rho^2\log\rho\,\bm y]
 =\rho^2\left[P(2)\log\rho+P'(2)\right]\bm y.
 \label{eq:frobenius-derivative}
\end{equation}
Writing the metric--scalar coefficients collectively as
\(\bm y_{(4)}^{\rm tot}\) and the logarithmic coefficients as
\(\bm y_{\log}\), the two resonant rows are
\begin{align}
 P(2)\bm y_{\log}&=0,
 \label{eq:weight-four-log-kernel}\\
 P(2)\bm y_{(4)}^{\rm tot}+P'(2)\bm y_{\log}
 +\bm R_{(4)}[g_{(0)},\phi_{(0)};g_{(2)},\phi_{(2)}]&=0,
 \label{eq:weight-four-finite-row}
\end{align}
where \(\bm R_{(4)}\) is completely local after substituting
Eq.~\eqref{eq:weight-two-solution}. If \(L\) spans the left kernel of
\(P(2)\), the Fredholm condition
\begin{equation}
 L\!\left[P'(2)\bm y_{\log}+\bm R_{(4)}\right]=0
 \label{eq:weight-four-solvability}
\end{equation}
fixes the logarithmic obstruction. The complementary projection fixes a
source-local particular solution, whereas
\(\bm y_{(4)}^{\rm state}\in\ker P(2)\) carries the normalizable state.
We find from Eq.~\eqref{eq:indicial-factors} that the tensor and scalar
response determinants govern the full Frobenius structure of the radial
system.

\section{Counterterms, Weyl anomaly, and logarithmic obstruction}
\label{sec:counterterms-anomaly}

The regulated on-shell functional combines the cutoff bulk integral with the Dirichlet surface functional,
\begin{equation}
 S_{\rm reg}(\varepsilon)=S_{\rm bulk}^{\rho\geq\varepsilon}
 +S_{\partial,D}^{\rho=\varepsilon}.
 \label{eq:regulated-action}
\end{equation}
Substitution of the solved FG data organizes it by derivative weight,
\begin{equation}
 S_{\rm reg}=S_{\rm fin}
 +\sum_{w<4}\varepsilon^{(w-4)/2}S_{(w)}
 +(\log\varepsilon)S_{(4)}+o(1).
 \label{eq:regulated-expansion}
\end{equation}
The counterterm descent removes the power and logarithmic coefficients:
\begin{equation}
 S_{\rm ct}=-\left[
 \sum_{w<4}\varepsilon^{(w-4)/2}S_{(w)}
 +(\log\varepsilon)S_{(4)}\right],
 \qquad
 S_{\rm ren}=\lim_{\varepsilon\to0}
 (S_{\rm reg}+S_{\rm ct}+S_{\rm fin}^{\rm scheme}).
 \label{eq:counterterm-descent}
\end{equation}
The radial Hamilton--Jacobi derivative expansion expresses the same hierarchy through
\begin{equation}
 S_{\rm loc}=\frac{1}{16\pi G_5}\int\dd^4x\sqrt{-\gamma}
 \left[\cU_{(0)}+\cU_{(2)}+(\log\varepsilon)\cU_{(4)}\right],
 \label{eq:hj-functional}
\end{equation}
where the canonical responses \(\pi^{ij}=\delta S_{\rm loc}/\delta\gamma_{ij}\) and \(\pi_\phi=\delta S_{\rm loc}/\delta\phi\) organize the radial constraint in ascending derivative weight~\cite{deBoerVerlindeVerlinde2000,MartelliMueck2003,Papadimitriou2010,Papadimitriou2011,ElvangHadjiantonis2016}. The completed cutoff currents supply the canonical data, and the explicit FG recursion fixes the exact SDLH coupling coefficients entering this descent.

\subsection{Power counterterms}
\label{subsec:power-counterterms}

The intrinsic weight-zero and weight-two counterterm density is
\begin{equation}
 \cB_{\rm ct}^{(0+2)}=c_0+c_R^{\rm ct}\widehat R+c_X\,X_kX^k,
 \label{eq:power-ct-density}
\end{equation}
with
\begin{equation}
 c_0=-\frac{2(3+\alpha_0s^2)}{3\ell},
 \qquad
 c_R^{\rm ct}=-\frac{9\ell^2+12\alpha_2s^2-8\alpha_3s^3-\alpha_0\ell^2s^2}{6\ell},
 \qquad
 c_X=\frac{3\alpha_2-2\alpha_3s}{\ell}.
 \label{eq:power-ct-coefficients}
\end{equation}
The corresponding cutoff currents are
\begin{align}
\cJ_{\rm ct}^{(0+2)i}{}_j
&=\frac{1}{2}c_0\delta^i_j-c_R^{\rm ct}\widehat\cG^i{}_j
+c_X\left(\frac{1}{2}X_kX^k\delta^i_j-X^iX_j\right),
\label{eq:power-ct-metric}\\
\cJ_{\rm ct}^{(0+2)\phi}&=-2c_X\widehat\Box\phi.
\label{eq:power-ct-scalar}
\end{align}
Equations~\eqref{eq:power-ct-coefficients} remove the \(\varepsilon^{-2}\) and \(\varepsilon^{-1}\) divergences and preserve one common metric--scalar source functional.

\subsection{Compact anomaly and its nine-density expansion}
\label{subsec:compact-anomaly}

Define the four-dimensional Euler and Weyl densities
\begin{equation}
 E_4=R_{ijkl}R^{ijkl}-4R_{ij}R^{ij}+R^2,
 \qquad C^2=C_{ijkl}C^{ijkl},
 \label{eq:euler-weyl}
\end{equation}
and the polynomial
\begin{align}
P_E={}&15\ell^4+132\alpha_2\ell^2s^2-8\alpha_0\ell^4s^2
-64\alpha_3\ell^2s^3+336\alpha_2^2s^4
-36\alpha_0\alpha_2\ell^2s^4+\alpha_0^2\ell^4s^4\notag\\
&-288\alpha_2\alpha_3s^5+16\alpha_0\alpha_3\ell^2s^5
+64\alpha_3^2s^6.
\label{eq:pe-polynomial}
\end{align}
The exact coefficients are
\begin{equation}
 a_E=-\frac{\ell P_E}{16\Delta_T},
 \qquad
 a_C=\frac{\ell U^2}{48\Delta_T},
 \qquad
 \kappa=\frac{N^2}{\ell\Delta_T\Delta_S}.
 \label{eq:anomaly-coefficients}
\end{equation}
The weight-four logarithmic row then condenses into
\begin{equation}
 \cA_{\rm reg}=a_EE_4+a_CC^2
 -\frac{\kappa}{4}\left(X_kX^k+2sH-\frac{2}{3}s^2R\right)^2.
 \label{eq:compact-anomaly}
\end{equation}
The scalar-source contribution is governed by the single Weyl-covariant combination
\begin{equation}
 \Xi\equiv X_kX^k+2s\Box_{(0)}\phi_{(0)}-\frac{2}{3}s^2R.
 \label{eq:xi-combination}
\end{equation}
Under the generalized Weyl transformation \eqref{eq:weyl-variations}, the elementary source jets transform as
\begin{align}
 \delta_\sigma X_kX^k&=-2\sigma X_kX^k-4s X^iD_i\sigma,
 \notag\\
 \delta_\sigma H&=-2\sigma H+2X^iD_i\sigma-2sD^2\sigma,
 \notag\\
 \delta_\sigma R&=-2\sigma R-6D^2\sigma .
 \label{eq:weyl-source-jets}
\end{align}
Their inhomogeneous terms cancel inside \(\Xi\), giving
\begin{equation}
 \delta_\sigma\Xi=-2\sigma\Xi,
 \qquad
 \delta_\sigma(\sqrt{-g_{(0)}}\,\Xi^2)=0.
 \label{eq:xi-weyl-covariance}
\end{equation}
Together with the Weyl invariance of \(\sqrt{-g_{(0)}}C^2\) and the Euler descent, Eq.~\eqref{eq:xi-weyl-covariance} gives the Wess--Zumino consistency relation
\begin{equation}
 [\delta_{\sigma_1},\delta_{\sigma_2}]S_{\rm ren}=0,
 \qquad
 \delta_{\sigma_1}\!\int\!\sqrt{-g_{(0)}}\,\sigma_2\cA_{\rm reg}
 -(1\leftrightarrow2)=0.
 \label{eq:wess-zumino-consistency}
\end{equation}
The square \(\Xi^2\) packages the boundary kinetic invariant, scalar Laplacian, boundary curvature, and radial beta spurion into one exact geometric density. Building on the covariant anomaly for arbitrary sources in~\cite{WuStone2026}, its compact cohomological representative \(\{E_4,C^2,\Xi^2\}\) and paired metric--scalar variations determine the logarithmic obstruction, finite canonical currents, and generalized trace identity in one source convention.

The integrated four-derivative basis is
\begin{equation}
 \big\{R^2,\ R_{ij}R^{ij},\ C^2,\ RX_kX^k,\ R_{ij}X^iX^j,
 (X_kX^k)^2,\ H^2,\ X_kX^kH,\ RH\big\}.
 \label{eq:nine-density-basis}
\end{equation}
In this order, Eq.~\eqref{eq:compact-anomaly} has coefficient vector
\begin{equation}
 \left\{
 \frac{2a_E}{3}-\frac{\kappa s^4}{9},
 -2a_E,
 a_E+a_C,
 \frac{\kappa s^2}{3},
 0,
 -\frac{\kappa}{4},
 -\kappa s^2,
 -\kappa s,
 \frac{2\kappa s^3}{3}
 \right\}.
 \label{eq:nine-density-coefficients}
\end{equation}
Equivalently, the basis conversion is the exact linear map
\begin{equation}
 \bm a_{9}=B_\Xi
 \begin{pmatrix}a_E\\a_C\\\kappa\end{pmatrix},
 \qquad
 B_\Xi=
 {\renewcommand{\arraystretch}{0.4}%
 \setlength{\arraycolsep}{12pt}%
 \begin{pmatrix}
  2/3 & 0 & -s^4/9\\
  -2      & 0 & 0\\
  1      & 1 & 0\\
  0      & 0 & s^2/3\\
  0      & 0 & 0\\
  0      & 0 & -1/4\\
  0      & 0 & -s^2\\
  0      & 0 & -s\\
  0      & 0 & 2s^3/3
 \end{pmatrix}},
 \label{eq:compact-to-nine-basis-map}
\end{equation}
where the row order is Eq.~\eqref{eq:nine-density-basis}. This map fixes every sign and normalization entering the paired metric--scalar variations.
The paired functional variations span eight independent directions. The integrated Euler density supplies the unique four-dimensional Lanczos null vector. The logarithmic counterterm is
\begin{equation}
 S_{\rm ct}^{\log}=-\frac{\log\varepsilon}{16\pi G_5}
 \int_{\rho=\varepsilon}\dd^4x\sqrt{-\gamma}\,
 \widehat\cA_{\rm reg}[\gamma,\phi].
 \label{eq:log-counterterm}
\end{equation}
At the cutoff, define \(\Xi_\gamma=X_{\gamma\,k}X_\gamma^k+2s\widehat\Box\phi-\frac23s^2\widehat R\) and construct \(\widehat\cA_{\rm reg}[\gamma,\phi]\) from \((\widehat E_4,\widehat C^2,\Xi_\gamma^2)\). The FG scaling gives
\begin{equation}
 \Xi_\gamma=\varepsilon\Xi+O(\varepsilon^2\log\varepsilon),
 \qquad
 \sqrt{-\gamma}\,\widehat\cA_{\rm reg}[\gamma,\phi]
 =\sqrt{-g_{(0)}}\,\cA_{\rm reg}[g_{(0)},\phi_{(0)}]
 +O(\varepsilon\log\varepsilon).
 \label{eq:cutoff-anomaly-uplift}
\end{equation}
Equation~\eqref{eq:cutoff-anomaly-uplift} is the explicit cutoff-to-source uplift used in Eq.~\eqref{eq:log-counterterm}.

\subsection{From anomaly variation to logarithmic obstruction}
\label{subsec:obstruction}

For every integrated density \(\cI_A\) in Eq.~\eqref{eq:nine-density-basis}, define paired Euler responses by
\begin{equation}
 \delta\int\dd^4x\sqrt{-g_{(0)}}\,\cI_A
 =\int\dd^4x\sqrt{-g_{(0)}}
 \left(\cV_A^{ij}\delta g_{(0)ij}+\cS_A\delta\phi_{(0)}\right).
 \label{eq:paired-variation}
\end{equation}
Let \(a_{{\rm reg},A}\) denote the ordered entries in Eq.~\eqref{eq:nine-density-coefficients}. The anomaly responses are
\begin{equation}
 V^i{}_j=\sum_{A=1}^9a_{{\rm reg},A}\cV_A{}^i{}_j,
 \qquad S=\sum_{A=1}^9a_{{\rm reg},A}\cS_A,
 \qquad (V^i{}_j,S)=-(V_{\rm ct}{}^i{}_j,S_{\rm ct}).
\label{eq:anomaly-response-definition}
\end{equation}
Define
\begin{equation}
 \mathfrak a=\alpha_0\ell^2+6\alpha_2-4\alpha_3s,
 \qquad
 \mathfrak b=3\alpha_2-2\alpha_3s,
 \qquad
 \mathfrak a\Delta_T-8s^2\mathfrak b^2=\ell^2\Delta_S .
 \label{eq:obstruction-block-identity}
\end{equation}
With \(V_{\rm TF}{}^i{}_j=V^i{}_j-\frac14\delta^i_j\tr V\) and
\(h_{(4)}\equiv h_{(4)i}{}^i\), the direct logarithmic response system is
\begin{align}
 V_{\rm TF}{}^i{}_j
 &= \frac{2\Delta_T}{3\ell^3}h_{(4){\rm TF}}{}^i{}_j,
 \label{eq:obstruction-direct-tf}\\
 \begin{pmatrix}\tr V\\ S\end{pmatrix}
 &=-\frac{2}{\ell^3}
 \begin{pmatrix}
 \Delta_T&16s\mathfrak b\\
 4s\mathfrak b&8\mathfrak a
 \end{pmatrix}
 \begin{pmatrix}h_{(4)}\\ \psi_{(4)}\end{pmatrix}.
 \label{eq:obstruction-direct-scalar}
\end{align}
The determinant of the displayed \(2\times2\) response operator, including
its common prefactor, is \(32\Delta_S/\ell^4\); hence it is invertible on
Eq.~\eqref{eq:regular-response-domain}. Its inverse yields
\begin{align}
h_{(4){\rm TF}}{}^i{}_j
&=\frac{3\ell^3}{2\Delta_T}V_{\rm TF}{}^i{}_j,
\label{eq:h4-tf}\\
h_{(4)}
&=\frac{\ell\left[-(\alpha_0\ell^2+6\alpha_2-4\alpha_3s)\tr V
+2s(3\alpha_2-2\alpha_3s)S\right]}{2\Delta_S},
\label{eq:h4-trace}\\
\psi_{(4)}
&=\frac{\ell\left[-\Delta_TS+4s(3\alpha_2-2\alpha_3s)\tr V\right]}{16\Delta_S}.
\label{eq:psi4}
\end{align}
Equations~\eqref{eq:h4-tf}--\eqref{eq:psi4} map the local Weyl anomaly into the metric and scalar logarithmic FG data. Thus the tensor and scalar response determinants control the anomaly coefficient, the Frobenius resonance, and the obstruction inversion.

\section{Finite stress tensor and scalar one-point function}
\label{sec:onepoints}

The renormalized generating functional is defined by the source convention in Eq.~\eqref{eq:intro-onepoint-definition}. The observable extraction begins with the finite coefficient operation
\begin{equation}
 \mathsf F[F]\equiv
 \operatorname{Coeff}_{\rho^2(\log\rho)^0}F(\rho).
 \label{eq:finite-operation}
\end{equation}
After inserting the exact weight-two solution and the logarithmic finiteness relations, define
\begin{align}
\cP^i{}_j&=\mathsf F\!\left[
\cJ_D{}^i{}_j+\cJ_{\rm ct}^{(0+2)i}{}_j\right],
\label{eq:pmetric}\\
\cP_\phi&=\mathsf F\!\left[
\cJ_D^\phi+\cJ_{\rm ct}^{(0+2)\phi}\right].
\label{eq:pscalar}
\end{align}
Equation~\eqref{eq:finite-operation} acts directly on the completed covariant currents in Eqs.~\eqref{eq:dirichlet-metric-current}, \eqref{eq:dirichlet-scalar-current}, \eqref{eq:power-ct-metric}, and \eqref{eq:power-ct-scalar}. After inserting the solved radial data it produces the exact source-current contributions as covariant local expressions in \(g_{(0)}\) and \(\phi_{(0)}\). Bianchi identities, derivative commutators, integrations by parts, and the Euler/Lanczos identity relate its equivalent termwise representatives.

The response-bearing weight-four block takes an especially compact form. Its exact mixing coefficients are
\begin{equation}
 C_g=-\frac{2\Delta_T}{3\ell^3},
 \qquad
 C_\phi=D_g=\frac{8s(3\alpha_2-2\alpha_3s)}{\ell^3},
 \qquad
 D_\phi=\frac{16(\alpha_0\ell^2+6\alpha_2-4\alpha_3s)}{\ell^3}.
 \label{eq:response-coefficients}
\end{equation}
They enter the total weight-four coefficients as
\begin{align}
\cP^i{}_j\big|_{\rm resp}
&=C_g\left(g_{(4)}^{{\rm tot},i}{}_j
-\delta^i_j\tr g_{(4)}^{\rm tot}\right)
+C_\phi\phi_{(4)}^{\rm tot}\delta^i_j,
\label{eq:metric-response}\\
\cP_\phi\big|_{\rm resp}
&=D_g\tr g_{(4)}^{\rm tot}+D_\phi\phi_{(4)}^{\rm tot}.
\label{eq:scalar-response}
\end{align}
Substitution of Eq.~\eqref{eq:weight-four-split} separates this block into its source-local and normalizable state contributions. The local summands combine with the remaining source-current terms generated by Eq.~\eqref{eq:finite-operation}; the state summands carry the normalizable data. The relation \(C_\phi=D_g\) expresses reciprocal metric--scalar mixing generated by the common renormalized functional.
Because both currents derive from the same renormalized functional, the mixed source Hessian is symmetric:
\begin{equation}
 \frac{\delta}{\delta\phi_{(0)}(y)}
 \left[\frac{\sqrt{-g_{(0)}(x)}}{2}\langle T^{ij}(x)\rangle\right]
 =
 \frac{\delta}{\delta g_{(0)ij}(x)}
 \left[\sqrt{-g_{(0)}(y)}\langle\cO_\phi(y)\rangle\right].
 \label{eq:mixed-hessian-integrability}
\end{equation}
Projecting Eq.~\eqref{eq:mixed-hessian-integrability} onto the total weight-four response gives \(C_\phi=D_g\), the reciprocity relation required by a single renormalized generating functional. The same functional Hessian combines this response block with the compact source-current remainder defined by Eq.~\eqref{eq:finite-operation}.

Let \((V_{\rm ct}{}^i{}_j,S_{\rm ct})\) denote the paired responses of the cancelling logarithmic density, and let \(f_A\) multiply the covariant weight-four finite densities in Eq.~\eqref{eq:nine-density-basis}. The complete observables in this scheme family are
\begin{equation}
\langle T^i{}_j\rangle=\frac{1}{8\pi G_5}
\left(\cP^i{}_j-\frac{1}{2}V_{\rm ct}{}^i{}_j
+\sum_Af_A\cV_A{}^i{}_j\right),
\label{eq:stress-tensor}
\end{equation}
\begin{equation}
\langle\cO_\phi\rangle=\frac{1}{16\pi G_5}
\left(\cP_\phi-\frac{1}{2}S_{\rm ct}
+\sum_Af_A\cS_A\right).
\label{eq:scalar-onepoint}
\end{equation}
Every logarithmic subtraction and finite scheme density enters through paired metric and scalar variations. Equations~\eqref{eq:stress-tensor} and \eqref{eq:scalar-onepoint}, together with the compact operation \eqref{eq:finite-operation}, therefore give the complete renormalized observables for arbitrary metric and scalar sources in a common convention.

An affine scalar source gives a compact nonhomogeneous probe. Set
\begin{equation}
 g_{(0)ij}=\eta_{ij},\qquad
 \phi_{(0)}=\phi_c+k_ix^i,\qquad
 X_i=k_i,\qquad H_{ij}=0.
 \label{eq:affine-source}
\end{equation}
The exact weight-two solution and anomaly become
\begin{align}
 g_{(2)ij}&=c_{XX}k_ik_j+c_{X^2}k^2\eta_{ij},
 &\phi_{(2)}&=b_{X^2}k^2,
 \notag\\
 \cA_{\rm reg}&=-\frac{\kappa}{4}(k^2)^2,
 &\partial_i\langle T^i{}_j\rangle&=\langle\cO_\phi\rangle k_j.
 \label{eq:affine-source-response}
\end{align}
This probe isolates scalar-gradient backreaction, the quartic scalar anomaly channel, and local momentum exchange in flat boundary geometry.

\section{Momentum constraint and Ward identities}
\label{sec:ward}

\subsection{Boundary diffeomorphisms and the mixed radial equation}

Under a boundary diffeomorphism generated by \(\xi^i\),
\begin{equation}
 \delta_\xi g_{(0)ij}=D_i\xi_j+D_j\xi_i,
 \qquad
 \delta_\xi\phi_{(0)}=\xi^jD_j\phi_{(0)}.
 \label{eq:diffeo-variations}
\end{equation}
Substitution into the source variation and one covariant integration by parts gives
\begin{equation}
 \delta_\xi S_{\rm ren}
 =-\int\dd^4x\sqrt{-g_{(0)}}\,\xi^j
 \left[D_i\langle T^i{}_j\rangle
 -\langle\cO_\phi\rangle D_j\phi_{(0)}\right].
 \label{eq:noether-variation}
\end{equation}
This source-side derivation identifies the current combination required by boundary covariance.

The bulk realization follows directly from diffeomorphism covariance. For each action sector \(a\), the variations
\begin{equation}
 \delta_\xi G_{MN}=2\nabla_{(M}\xi_{N)},
 \qquad
 \delta_\xi\phi=\xi^M\nabla_M\phi
\end{equation}
and one covariant integration by parts give
\begin{equation}
 -2\nabla^M\cE^{(a)}_{MN}+2\cE^{(a)}_\phi\nabla_N\phi=0.
 \label{eq:bulk-noether-identity}
\end{equation}
The second factor of two follows from the scalar normalization in
Eq.~\eqref{eq:scalar-eom-normalization}. Summing the six sectors and
projecting this identity onto the radial Gauss--Codazzi slicing yields the
finite-cutoff momentum relation
\begin{equation}
 -2D_i\cJ^i{}_j+\cJ^\phi D_j\phi
 =-\frac{4\rho}{\ell}\cE_{\rho j}.
\label{eq:finite-cutoff-momentum-noether}
\end{equation}
Here the un-subscripted currents include the intrinsic counterterm currents,
\begin{equation}
 \cJ^i{}_j=\cJ_D{}^i{}_j+\cJ_{\rm ct}{}^i{}_j,
 \qquad
 \cJ^\phi=\cJ_D^\phi+\cJ_{\rm ct}^\phi .
 \label{eq:finite-cutoff-total-currents}
\end{equation}
Each intrinsic counterterm functional satisfies its own homogeneous boundary
Noether identity, so its addition leaves the bulk right-hand side of
Eq.~\eqref{eq:finite-cutoff-momentum-noether} unchanged.
At vector weight five define
\begin{equation}
 \overline\cE_{\rho j}^{\rm nonlog}\equiv\overline\cE_{\rho j}^{[1,0]},
 \qquad
 \overline\cE_{\rho j}^{\log}\equiv\overline\cE_{\rho j}^{[1,1]} .
 \label{eq:momentum-coefficient-definition}
\end{equation}
The linear Laurent--log projection of Eq.~\eqref{eq:finite-cutoff-momentum-noether} gives
\begin{equation}
D_i\langle T^i{}_j\rangle
-\langle\cO_\phi\rangle D_j\phi_{(0)}
=\frac{\overline\cE_{\rho j}^{\rm nonlog}
-\tfrac12\overline\cE_{\rho j}^{\log}}{4\pi G_5\ell}.
\label{eq:diffeomorphism-ward}
\end{equation}
Imposing the mixed radial equation sets its right-hand side to zero. Its four components describe local energy--momentum transfer for arbitrary curved \(g_{(0)ij}(x)\) and local \(\phi_{(0)}(x)\). The factor \(1/2\) is the finite residue of \(\partial_\rho(\rho^2\log\rho)=2\rho\log\rho+\rho\)~\cite{MartelliMueck2003,Papadimitriou2011,CaceresEtAl2024}.

The boundary Noether current and the mixed radial equation are two projections of the same covariant variational identity.

\subsection{Radial Weyl transformation and generalized trace}

The logarithmic scalar profile changes the radial realization of a boundary
Weyl transformation. The Penrose--Brown--Henneaux radial component is
\(\xi^\rho=-2\sigma\rho+O(\rho^2)\), so
\(\delta_\sigma(s\log\rho)=\xi^\rho\partial_\rho(s\log\rho)=-2s\sigma\).
The compensating tangential component supplies the metric transformation. In
the present convention,
\begin{equation}
 \delta_\sigma g_{(0)ij}=2\sigma g_{(0)ij},
 \qquad
 \delta_\sigma\phi_{(0)}=-2s\sigma .
 \label{eq:weyl-variations}
\end{equation}
Applying Eq.~\eqref{eq:weyl-variations} to the common renormalized functional gives the scheme-resolved identity
\begin{equation}
\langle T^i{}_i\rangle-2s\langle\cO_\phi\rangle
=\frac{1}{8\pi G_5}\left[-\cA_{\rm reg}
+\sum_Af_A\left(\tr\cV_A-s\cS_A\right)\right].
\label{eq:trace-ward-scheme}
\end{equation}
The minimal finite scheme \(f_A=0\) selects the compact form
\begin{equation}
\langle T^i{}_i\rangle-2s\langle\cO_\phi\rangle
=-\frac{\cA_{\rm reg}}{8\pi G_5}.
\label{eq:trace-ward}
\end{equation}
The factor \(2s\) is the beta-spurion weight generated by the radial running of \(\phi\). The first term on the right-hand side is the full anomaly for arbitrary metric and scalar sources in Eq.~\eqref{eq:compact-anomaly}, and the second records the Weyl variation of finite local densities. Equations~\eqref{eq:diffeomorphism-ward} and \eqref{eq:trace-ward-scheme} place the momentum and dilatation constraints inside one boundary Noether algebra.

\subsection{Homogeneous reduction}
\label{subsec:flat-specialization}

Set \(g_{(0)ij}=\eta_{ij}\), choose constant \(\phi_{(0)}\), and take constant state data. Curvature and local source jets then vanish. With \(t_4=\tr g_{(4)}^{\rm state}\), the observable pair becomes
\begin{align}
\langle T^i{}_j\rangle_{\rm flat}
&=\frac{1}{8\pi G_5}\left[
C_g\left(g_{(4)}^{{\rm state},i}{}_j-\delta^i_jt_4\right)
+C_\phi\phi_{(4)}^{\rm state}\delta^i_j\right],
\label{eq:flat-t}\\
\langle\cO_\phi\rangle_{\rm flat}
&=\frac{1}{16\pi G_5}\left(D_gt_4+D_\phi\phi_{(4)}^{\rm state}\right).
\label{eq:flat-o}
\end{align}
The normalizable equation gives
\begin{equation}
8\alpha_0s\phi_{(4)}^{\rm state}
+(\alpha_0s^2-3)t_4=0,
\label{eq:flat-normalizable}
\end{equation}
and Eq.~\eqref{eq:trace-ward} reduces to
\begin{equation}
\langle T^i{}_i\rangle_{\rm flat}
-2s\langle\cO_\phi\rangle_{\rm flat}=0.
\label{eq:flat-trace}
\end{equation}
The homogeneous formulas arise by functionally differentiating the generating functional for arbitrary metric and scalar sources and then imposing \(g_{(0)ij}=\eta_{ij}\) and constant \(\phi_{(0)}\). On this configuration the source-local curvature and gradient currents, the covariant anomaly density, and the local momentum-exchange terms vanish, while the normalizable state response in Eqs.~\eqref{eq:flat-t} and \eqref{eq:flat-o} remains. Equation~\eqref{eq:flat-trace} is the homogeneous specialization of the generalized trace identity.

\section{The string-to-observable map}
\label{sec:implications}

We obtain a sequence of exact maps. Equation~\eqref{eq:weight-two-solution} maps boundary curvature and scalar jets to the local radial response. The same determinants enter the indicial equation \eqref{eq:indicial-factors}, the anomaly coefficients \eqref{eq:anomaly-coefficients}, the obstruction map \eqref{eq:h4-tf}--\eqref{eq:psi4}, and the normalizable mixing \eqref{eq:response-coefficients}. Hence \((\Delta_T,\Delta_S)\) organizes the near-boundary response from the first local backreaction to the finite observables. Their zero sets define the tensor and scalar Frobenius-reorganization surfaces.

The approach to these hypersurfaces exposes the exact poles of the regular Frobenius chart. For nonzero dimensionless distances \(\epsilon_T=\Delta_T/\ell^2\) and \(\epsilon_S=\Delta_S/\ell^2\), the transverse logarithmic response scales as \(h_{(4){\rm TF}}\sim\epsilon_T^{-1}\), the scalar/trace obstruction scales as \((h_{(4)},\psi_{(4)})\sim\epsilon_S^{-1}\), and the scalar-source anomaly channel scales as \(\kappa\sim(\epsilon_T\epsilon_S)^{-1}\). Parameter derivatives in \(J^{\rm HR}\) acquire the corresponding next inverse power. The determinant surfaces therefore locate the indicial-degeneracy strata and supply their tangent and normal coordinates directly through the holographic Jacobian.

The anomaly square \(\Xi^2\) supplies a second exact organization of the response. It combines the source kinetic invariant, scalar Laplacian, and Ricci scalar in the combination selected by the linear radial profile. Its functional derivatives give the metric and scalar obstruction sources, which the response matrix maps to \((h_{(4)ij},\psi_{(4)})\). The Weyl anomaly therefore maps source geometry to logarithmic bulk data, while the finite currents map the same data to \(\langle T_{ij}\rangle\) and \(\langle\cO_\phi\rangle\).

The full holographic map is
\begin{equation}
 \{\alpha_0,\alpha_2,\alpha_3;\ell,s\}_{V_1=V_2=0}
 \longmapsto
 \{a_E,a_C,\kappa;\Delta_T,\Delta_S;
 C_g,C_\phi,D_g,D_\phi;
 \langle T_{ij}\rangle,\langle\cO_\phi\rangle\}.
 \label{eq:observable-map}
\end{equation}
Composing Eq.~\eqref{eq:observable-map} with the explicit reduction map \eqref{eq:explicit-kk-coupling-map} creates a direct coefficient-frame-to-boundary map. The exact pullback is defined by
\begin{equation}
 J^{\rm red}_{aI}=
 \frac{\partial\cH_a}{\partial\cI_A}
 \frac{\partial\cI_A}{\partial\Theta_{\rm str}^I},
 \qquad
 \ker J^{\rm red}=\left\{\delta\Theta_{\rm str}^I:
 J^{\rm red}_{aI}\delta\Theta_{\rm str}^I=0\right\}.
 \label{eq:reduction-holographic-jacobian}
\end{equation}

\subsection{Closed inverse and string-coefficient identifiability}
\label{subsec:closed-string-inverse}

On the invariant chart, we evaluate the pullback in Eq.~\eqref{eq:reduction-holographic-jacobian} globally in closed form. Write
\begin{equation}
 x=\mathsf c_0,\qquad y=\mathsf c_2,\qquad z=\mathsf c_3,
 \qquad m=3y-2z,
 \label{eq:inverse-invariant-chart}
\end{equation}
and introduce the dimensionless branch polynomials
\begin{equation}
 T=3-x+4m,\qquad
 Q=3+24y-x-8z=T+12y,
 \qquad S=2xm-x^2+6m+3x.
 \label{eq:inverse-branch-polynomials}
\end{equation}
The vacuum equations and the definitions above give
\begin{equation}
 \Delta_T=\ell^2T,\qquad
 U=\ell^2Q,\qquad
 \Delta_S=\frac{\ell^2}{ s^2}S,
 \label{eq:inverse-branch-data}
\end{equation}
where \(U\) is the polynomial defined in Eq.~\eqref{eq:u-polynomial} and
entering \(a_C\). A minimal normalization-invariant
holographic fingerprint is
\begin{equation}
 G=C_g,\qquad A=sC_\phi=sD_g,\qquad
 B=s^2D_\phi,\qquad E=a_E+a_C.
 \label{eq:inverse-fingerprint}
\end{equation}
Direct substitution of Eqs.~\eqref{eq:anomaly-coefficients} and
\eqref{eq:response-coefficients} yields the exact forward map
\begin{equation}
 G=-\frac{2T}{3\ell},\qquad
 A=\frac{8m}{\ell},\qquad
 B=\frac{16(x+2m)}{\ell},\qquad
 E=-\frac{\ell^3}{48}(Q+T+6).
 \label{eq:inverse-forward-map}
\end{equation}
It admits the global rational inverse
\begin{align}
 \ell&=\frac{48}{ B-12A-24G},
 &m&=\frac{\ell A}{8},
 &x&=\frac{\ell(B-4A)}{16},
 \notag\\
 T&=-\frac{3\ell G}{2},
 &Q&=-\frac{48E}{\ell^3}-T-6,
 &y&=\frac{Q-T}{12},
 \notag\\
 &&z&=\frac{Q-T}{8}-\frac{m}{2}.
 \label{eq:inverse-global-rational}
\end{align}
The five-dimensional couplings then follow from
\begin{equation}
 \alpha_0=\frac{x}{ s^2},\qquad
 \alpha_2=\frac{\ell^2y}{ s^2},\qquad
 \alpha_3=\frac{\ell^2z}{ s^3},\qquad
 \alpha_4=\frac{\ell^2(-x-6y+8z)}{8s^4}.
 \label{eq:inverse-five-dimensional-couplings}
\end{equation}
For the fixed normalization \(\alpha_0=-4/5\), this inverse also supplies
the exact observable-slice condition
\begin{equation}
 s^2=-\frac{5x}{4}=-\frac{5\ell}{64}(B-4A),
 \qquad B-4A=-\frac{64s^2}{5\ell}.
 \label{eq:inverse-fixed-normalization-slice}
\end{equation}
Thus a real nonzero slope with \(\ell>0\) requires \(x<0\) and
\(B-4A<0\). The anomaly coefficients contain an especially direct inverse:
\begin{equation}
 a_E+a_C=-\frac{\ell^3}{2}\mathsf c_1
 =-\frac{\ell}{2}\alpha_1,
 \qquad
 \widetilde\alpha_{\rm GB}=\alpha_1
 =-\frac{2(a_E+a_C)}{\ell}.
 \label{eq:inverse-gauss-bonnet}
\end{equation}
This linear relation fixes the sign uniquely, whereas \(a_C\propto Q^2\)
alone leaves a twofold ambiguity.

We establish global identifiability on the regular finite-radius chart:
\begin{equation}
 \det\frac{\partial(G,A,B,E)}{\partial(\ell,x,y,z)}=\frac{128}{\ell},
 \qquad
 \rank J^{\rm HR}_{/\mathrm{norm}}=4,
 \qquad
 \ker J^{\rm HR}_{/\mathrm{norm}}=\{0\}.
 \label{eq:inverse-hr-rank}
\end{equation}
The response-only block \((G,A,B)\) has the one-dimensional kernel
\begin{equation}
 v_R=(0,0,2,3)_{(\ell,x,y,z)},
 \qquad \dd E(v_R)=-\frac{\ell^3}{2}\ne0,
 \label{eq:inverse-response-kernel}
\end{equation}
so the curvature-anomaly sum is precisely the channel that removes this
degeneracy. On the fixed \(\alpha_0=-4/5\) chart,
\begin{equation}
 \det\frac{\partial(G,A,B,E)}{\partial(\ell,s,y,z)}
 =-\frac{1024s}{5\ell},
 \label{eq:inverse-fixed-normalization-rank}
\end{equation}
which is nonzero for \(s\ell\ne0\).

At fixed integer \(n\), setting \(\beta=2/n\) in
Eq.~\eqref{eq:explicit-kk-coupling-map} gives the explicit string-frame
inverse
\begin{align}
 \widetilde\alpha_{\rm GB}&=\alpha_1,
 &a_1&=\frac{\alpha_2}{\alpha_1}+\frac{16(n-1)}{ n},
 \notag\\
 a_3&=\frac{\alpha_3}{\alpha_1}-3a_1
 +\frac{16(n-1)(n-2)}{ n^2},
 \notag\\
 a_4&=\frac{\alpha_4}{\alpha_1}
 -\frac{16(n-1)^2(n-2)}{ n^3}
 +\frac{4a_1(n-3)}{ n^2}.
 \label{eq:inverse-string-frame}
\end{align}
The reduction Jacobian and the vacuum-branch Jacobian factorize as
\begin{equation}
 \det\frac{\partial(\alpha_1,\alpha_2,\alpha_3,\alpha_4)
 }{\partial(\widetilde\alpha_{\rm GB},a_1,a_3,a_4)}
 =\widetilde\alpha_{\rm GB}^{3},
 \qquad
 \det\frac{\partial(V_1,V_2)}{\partial(\ell,s)}
 =\frac{8\ell\Delta_S}{ s}.
 \label{eq:inverse-factorized-jacobians}
\end{equation}
Consequently,
\begin{equation}
 \det\frac{\partial(G,A,B,E)
 }{\partial(\widetilde\alpha_{\rm GB},a_1,a_3,a_4)}
 =-\frac{12288s^9\widetilde\alpha_{\rm GB}^{3}
 }{5\ell^6\Delta_S},
 \qquad
 \rank J^{\rm red}=4,\qquad \ker J^{\rm red}=\{0\},
 \label{eq:inverse-reduction-rank}
\end{equation}
on \(s\Delta_T\Delta_S\widetilde\alpha_{\rm GB}\ne0\). The integer \(n\)
labels discrete compactification hypotheses, each carrying the full-rank
continuous inverse.

The remaining holographic channels obey the independent algebraic identities
\begin{equation}
 \frac{48a_CT}{\ell^3}=Q^2,\qquad
 \frac{s^4\kappa}{\ell^3}
 =\frac{[T^2+(x-3)Q]^2}{64TS},
 \qquad C_\phi=D_g.
 \label{eq:inverse-redundant-checks}
\end{equation}
The finite boundary loci of this chart are
\(s=0\), \(\Delta_T=0\), \(\Delta_S=0\),
\(\widetilde\alpha_{\rm GB}=0\), and \(\ell=0\); the non-finite boundary
\(\ell\to\infty\) is equivalent to \(B-12A-24G\to0\).

Black-object and propagation calculations in the same invariant coordinates
furnish an independent gravitational block \(J^{\rm grav}\)~\cite{AnabalonCisternaOliva2014,LIGO2017,EzquiagaZumalacarregui2017}.
The holographic inverse resolves every continuous coupling direction at a
generic fixed-\(n\) point, whereas the gravitational block probes global
branches, horizons, and discrete compactification hypotheses.

Our results also clarify the role of higher-curvature and scalar--tensor structures in holography. The coefficients \(a_E\) and \(a_C\) measure curvature response, \(\kappa\) measures the scalar-source anomaly channel, and \(C_g,C_\phi,D_g,D_\phi\) form the state-response matrix; Eq.~\eqref{eq:diffeomorphism-ward} governs local momentum exchange. Differentiating or evaluating the same renormalized functional yields correlation functions with spacetime-dependent sources, thermal states, black branes, and transport, all with exact coupling dependence.

\section{Conclusions}
\label{sec:conclusions}

We have established a complete holographic dictionary for the five-dimensional SDLH theory on its exact linear-dilaton AlAdS\(_5\) branch, with an arbitrary boundary metric and spacetime-dependent scalar source. We separate local backreaction, logarithmic obstruction, and normalizable state response through the boundary-covariant radial hierarchy, and derive the stress tensor and scalar one-point function from the finite canonical currents.

We compute the anomaly in the cohomological form \(a_EE_4+a_CC^2-\kappa\Xi^2/4\), whose paired variations generate the metric and scalar logarithmic coefficients. We show that the same renormalized functional enforces reciprocal tensor--scalar mixing and yields the source-dependent momentum and trace Ward identities. The determinants \(\Delta_T\) and \(\Delta_S\) govern the local recursion, the Frobenius spectrum, the anomaly channel, and the finite response matrix.

Furthermore, we obtain a global rational inverse for the invariant quartet \((C_g,sC_\phi,s^2D_\phi,a_E+a_C)\) on the regular branch. Its Jacobian has maximal rank and no continuous kernel at fixed compactification dimension. In particular, \(a_E+a_C=-\ell\alpha_1/2\) reconstructs the higher-dimensional Gauss--Bonnet coefficient directly, while the remaining response and anomaly channels overdetermine the inverse through independent algebraic consistency relations.

Our results connect the heterotic coefficient frame and Kaluza--Klein reduction directly to boundary curvature response, scalar insertions, normalizable state data, and local momentum exchange. The exact holographic inverse supplies the ultraviolet block of a joint compactification and gravitational-observable reconstruction, establishing the SDLH theory as a multi-channel probe of string-selected scalar--tensor dynamics.

\appendix

\section{Conventions and radial tensor identities}
\label{app:conventions-radial}

We use Lorentzian signature $(-++++)$ and the curvature convention
\begin{equation}
 [\nabla_M,\nabla_N]V^P=R^P{}_{QMN}V^Q,
 \qquad R_{QN}=R^P{}_{QPN},
 \qquad \Box=G^{MN}\nabla_M\nabla_N .
 \label{eq:app-curvature-convention}
\end{equation}
On the AdS branch this gives
\(R_{MNRS}=-\ell^{-2}(G_{MR}G_{NS}-G_{MS}G_{NR})\).
Bulk indices $M,N,\ldots$ run over the five-dimensional spacetime, and
boundary indices $i,j,\ldots$ run over the four-dimensional cutoff slices.
The bulk metric is $G_{MN}$, the induced cutoff metric is $\gamma_{ij}$, and
the boundary source metric is $g_{(0)ij}$. We use
\begin{align}
 X&=G^{MN}\nabla_M\phi\nabla_N\phi,
 &\mathcal G_{MN}&=R_{MN}-\frac12G_{MN}R,\notag\\
 \mathcal L_{\rm GB}&=R_{MNRS}R^{MNRS}-4R_{MN}R^{MN}+R^2.
 \label{eq:app-bulk-definitions}
\end{align}
The source convention reads
\begin{equation}
 \delta S_{\rm ren}=\int\!\mathrm d^4x\sqrt{-g_{(0)}}
 \left[
 \frac12\langle T^{ij}\rangle\delta g_{(0)ij}
 +\langle\mathcal O_\phi\rangle\delta\phi_{(0)}
 \right].
 \label{eq:app-source-convention}
\end{equation}
It produces the diffeomorphism combination
$D_i\langle T^i{}_j\rangle-\langle\mathcal O_\phi\rangle
D_j\phi_{(0)}$ and the radial-Weyl combination
$\langle T^i{}_i\rangle-2s\langle\mathcal O_\phi\rangle$.

\subsection{FG slicing and extrinsic geometry}

The Fefferman--Graham slicing is
\begin{equation}
 \mathrm ds^2=\frac{\ell^2}{4\rho^2}\mathrm d\rho^2
 +\rho^{-1}g_{ij}(\rho,x)\mathrm dx^i\mathrm dx^j,
 \qquad
 N=\frac{\ell}{2\rho},
 \qquad N^i=0.
 \label{eq:app-fg-metric}
\end{equation}
At the small-$\rho$ regulator, the outward spacelike normal and extrinsic
curvature are
\begin{equation}
 n^2=+1,
 \qquad
 n_{\rm out}=-\frac{2\rho}{\ell}\partial_\rho,
 \qquad
 K_{ij}=\frac12\mathcal L_n\gamma_{ij},
 \qquad
 K_{ij}^{\rm out}=\frac1\ell\left(\rho^{-1}g_{ij}-g'_{ij}\right).
 \label{eq:app-normal-extrinsic}
\end{equation}
The scalar data on each slice are
\begin{equation}
 v=n^M\nabla_M\phi,
 \qquad
 X_i=D_i\phi,
 \qquad
 H_{ij}=D_iD_j\phi,
 \qquad H=D^2\phi.
 \label{eq:app-cutoff-scalar-data}
\end{equation}

The Gauss--Codazzi decomposition gives
\begin{align}
{}^{(5)}R_{ijkl}
&=\widehat R_{ijkl}-K_{ik}K_{jl}+K_{il}K_{jk},
\label{eq:app-gauss}\\
{}^{(5)}R_{\perp ijk}
&=D_jK_{ik}-D_kK_{ij},
\label{eq:app-codazzi}\\
\mathcal G_{\perp i}
&=D_jK^j{}_i-D_iK,
\label{eq:app-momentum-geometry}\\
{}^{(5)}R
&=\widehat R+K^2-K_{ij}K^{ij}
-2\nabla_M(Kn^M-a^M),
\qquad a_i=D_i\log N=0.
\label{eq:app-scalar-gauss}
\end{align}
The completed Dirichlet density is
\begin{equation}
 \mathcal B_D=2K+4\alpha_1\left(J-2\widehat{\mathcal G}_{ij}K^{ij}\right)
 +\alpha_2\left(K^{ij}X_iX_j-KX_kX^k\right)
 -\alpha_3v\left(X_kX^k+\frac13v^2\right),
 \label{eq:app-dirichlet-density}
\end{equation}
where
\begin{equation}
 J_{ij}=\frac13\left(
 2KK_{ik}K^k{}_j+K_{kl}K^{kl}K_{ij}
 -2K_{ik}K^{kl}K_{lj}-K^2K_{ij}
 \right),
 \qquad J=J^i{}_i.
 \label{eq:app-myers-tensor}
\end{equation}
The intrinsic double-dual tensor used in the metric current is
\begin{align}
\widehat P_{ikjl}={}&\widehat R_{ikjl}
+\widehat R_{kj}\gamma_{il}-\widehat R_{kl}\gamma_{ij}
-\widehat R_{ij}\gamma_{kl}+\widehat R_{il}\gamma_{kj}
\notag\\
&+\frac12\widehat R
(\gamma_{ij}\gamma_{kl}-\gamma_{il}\gamma_{kj}).
\label{eq:app-double-dual}
\end{align}

\subsection{Exact FG component identities}

Set
\begin{equation}
 A^i{}_j=(g^{-1}g')^i{}_j,
 \qquad {}'\equiv\partial_\rho.
 \label{eq:app-a-matrix}
\end{equation}
Direct evaluation in the gauge \eqref{eq:app-fg-metric} gives
\begin{align}
R_{\rho i}
&=\frac12\left(D^jg'_{ij}-D_i\operatorname{tr}A\right),
\label{eq:app-rho-i}\\
R_{\rho\rho}
&=-\rho^{-2}-\frac12\operatorname{tr}(g^{-1}g'')
+\frac14\operatorname{tr}(A^2),
\label{eq:app-rho-rho}\\
\Box_5\phi
&=\frac4{\ell^2}\left[
\rho^2\phi''-\rho\phi'
+\frac{\rho^2}{2}\operatorname{tr}A\,\phi'
\right]+\rho\Box_g\phi,
\label{eq:app-box-fg}\\
X
&=\frac{4\rho^2}{\ell^2}(\phi')^2
+\rho g^{ij}D_i\phi D_j\phi,
\label{eq:app-x-fg}\\
\sqrt{-G}
&=\frac\ell2\rho^{-3}\sqrt{-g}.
\label{eq:app-det-fg}
\end{align}
The mixed component \eqref{eq:app-rho-i} carries one free boundary index and
generates the four radial momentum constraints.

\subsection{Laurent--log coefficient algebra}

For
\begin{equation}
 F(\rho)=\sum_{p,q}F_{p,q}\rho^p(\log\rho)^q,
 \label{eq:app-formal-series}
\end{equation}
the radial derivative and product laws are
\begin{align}
\partial_\rho\left[\rho^p(\log\rho)^q\right]
&=p\rho^{p-1}(\log\rho)^q
+q\rho^{p-1}(\log\rho)^{q-1},
\label{eq:app-log-derivative}\\
(FG)_{p,q}
&=\sum_{\substack{r+s=p\\m+n=q}}F_{r,m}G_{s,n}.
\label{eq:app-series-product}
\end{align}
For $B=g^{-1}$,
\begin{equation}
 B_{0,0}=g_{(0)}^{-1},
 \qquad
 B_{p,q}=-g_{(0)}^{-1}
 \sum_{(r,m)\neq(0,0)}g_{r,m}B_{p-r,q-m},
 \label{eq:app-inverse-recursion}
\end{equation}
and
\begin{equation}
 \partial_\rho\log\sqrt{-g}
 =\frac12\operatorname{tr}(g^{-1}g').
 \label{eq:app-determinant-recursion}
\end{equation}
The normalized equation blocks are
\begin{equation}
 \overline{\mathcal E}_{\rho\rho}=\rho^2\mathcal E_{\rho\rho},\qquad
 \overline{\mathcal E}^{i}{}_{j}=\rho g^{ik}\mathcal E_{kj},\qquad
 \overline{\mathcal E}_{\rho i}=\mathcal E_{\rho i},\qquad
 \overline{\mathcal E}_{\phi}=\mathcal E_{\phi}.
 \label{eq:app-normalized-eom-blocks}
\end{equation}
Their coefficients are
\begin{equation}
 \overline{\mathcal E}_{A}^{[p,q]}
 \equiv\operatorname{Coeff}_{\rho^p(\log\rho)^q}\overline{\mathcal E}_{A},
 \qquad
 w=2p\quad(A=\rho\rho,i{\,}^{j},\phi),
 \qquad w=2p+3\quad(A=\rho i).
 \label{eq:app-eom-coefficients}
\end{equation}
These definitions generate the leading, weight-two, weight-four logarithmic,
weight-four finite, and state-constraint rows in one ordered algebra.

\section{Paired variations of the weight-four density basis}
\label{app:paired-variations}

For each density $\mathcal I_A$ in Eq.~\eqref{eq:nine-density-basis}, define
the paired Euler derivatives by
\begin{equation}
 \delta\int\!\mathrm d^4x\sqrt{-g_{(0)}}\,\mathcal I_A
 =\int\!\mathrm d^4x\sqrt{-g_{(0)}}
 \left(\mathcal V_A^{ij}\delta g_{(0)ij}
 +\mathcal S_A\delta\phi_{(0)}\right).
 \label{eq:app-paired-variation-definition}
\end{equation}
Every curvature tensor and covariant derivative below belongs to
$g_{(0)ij}$.

\subsection{\texorpdfstring{$R^2$}{R-squared}}

\begin{align}
\mathcal V_{R^2}{}^i{}_j
&=-2RR^i{}_j+\frac12\delta^i_jR^2
+2D^iD_jR-2\delta^i_jD^2R,
&\mathcal S_{R^2}&=0.
\label{eq:app-var-r2}
\end{align}

\subsection{\texorpdfstring{$R_{kl}R^{kl}$}{Ricci-squared}}

\begin{align}
\mathcal V_{R_{kl}^2}{}^i{}_j
&=\frac12\delta^i_jR_{kl}R^{kl}
-2R^{kl}R^i{}_{kjl}+D^iD_jR-D^2R^i{}_j
-\frac12\delta^i_jD^2R,
&\mathcal S_{R_{kl}^2}&=0.
\label{eq:app-var-ricci2}
\end{align}

\subsection{\texorpdfstring{$C_{klmn}C^{klmn}$}{Weyl-squared}}

\begin{equation}
 \mathcal V_{C^2}{}^i{}_j
 =2\mathcal V_{R_{kl}^2}{}^i{}_j
 -\frac23\mathcal V_{R^2}{}^i{}_j,
 \qquad
 \mathcal S_{C^2}=0.
 \label{eq:app-var-weyl2}
\end{equation}

\subsection{\texorpdfstring{$RX_kX^k$}{R times X-squared}}

\begin{align}
\mathcal V_{RX_kX^k}{}^i{}_j={}&-RX^iX_j-R^i{}_jX_kX^k
+\frac12\delta^i_jRX_kX^k+2X^kD_kH^i{}_j+2H^i{}_kH^k{}_j
\notag\\
&-2\delta^i_j(X^kD^2X_k+H_{kl}H^{kl})
+2R^i{}_{kjl}X^kX^l,
\label{eq:app-var-rq2-metric}\\
\mathcal S_{RX_kX^k}={}&-2RH-2(D_iR)X^i.
\label{eq:app-var-rq2-scalar}
\end{align}

\subsection{\texorpdfstring{$R_{kl}X^kX^l$}{Ricci XX}}

\begin{align}
\mathcal V_{R_{kl}X^kX^l}{}^i{}_j={}&H^i{}_jH-R_{jk}X^iX^k
-R^i{}_kX_jX^k+X^kD_kH^i{}_j
-\frac12\delta^i_jH^2
\notag\\
&-\delta^i_jX^kD^2X_k+\delta^i_jR_{kl}X^kX^l
-\frac12\delta^i_jH_{kl}H^{kl},
\label{eq:app-var-ricciqq-metric}\\
\mathcal S_{R_{kl}X^kX^l}={}&-(D_iR)X^i-2R_{ij}H^{ij}.
\label{eq:app-var-ricciqq-scalar}
\end{align}

\subsection{\texorpdfstring{$(X_kX^k)^2$}{X-fourth}}

\begin{align}
\mathcal V_{(X_kX^k)^2}{}^i{}_j
&=-2X_kX^kX^iX_j+\frac12\delta^i_j(X_kX^k)^2,
&\mathcal S_{(X_kX^k)^2}
&=-4X_kX^kH-8X^iX^jH_{ij}.
\label{eq:app-var-q4}
\end{align}

\subsection{\texorpdfstring{$H^2$}{H-squared}}

\begin{align}
\mathcal V_{H^2}{}^i{}_j={}&X_jD^2X^i+X^iD^2X_j
-R_{jk}X^iX^k-R^i{}_kX_jX^k
-\frac12\delta^i_jH^2
\notag\\
&-\delta^i_jX^kD^2X_k+\delta^i_jR_{kl}X^kX^l,
\label{eq:app-var-h2-metric}\\
\mathcal S_{H^2}={}&2D^2H.
\label{eq:app-var-h2-scalar}
\end{align}

\subsection{\texorpdfstring{$X_kX^kH$}{X-squared H}}

\begin{align}
\mathcal V_{X_kX^kH}{}^i{}_j
&=-X^iX_jH+X_jH^i{}_kX^k+X^iH_{jk}X^k
-\delta^i_jX^kH_{kl}X^l,
\label{eq:app-var-q2h-metric}\\
\mathcal S_{X_kX^kH}
&=-2H^2+2R_{ij}X^iX^j+2H_{ij}H^{ij}.
\label{eq:app-var-q2h-scalar}
\end{align}

\subsection{\texorpdfstring{$RH$}{R H}}

\begin{align}
\mathcal V_{RH}{}^i{}_j={}&\frac12(D^iR)X_j
+\frac12X^iD_jR-R^i{}_jH+D^2H^i{}_j
-(D^iR_{jk})X^k-(D_jR^i{}_k)X^k
\notag\\
&+(D_kR^i{}_j)X^k-\frac12\delta^i_j(D_kR)X^k
-R_{jk}H^{ik}-R^i{}_kH^k{}_j
-\delta^i_jD^2H+2R^i{}_{kjl}H^{kl},
\label{eq:app-var-rh-metric}\\
\mathcal S_{RH}={}&D^2R.
\label{eq:app-var-rh-scalar}
\end{align}

The Euler combination satisfies
\begin{equation}
 \mathcal V_{C^2}{}^i{}_j
 -2\mathcal V_{R_{kl}^2}{}^i{}_j
 +\frac23\mathcal V_{R^2}{}^i{}_j=0.
 \label{eq:app-euler-null}
\end{equation}
Consequently the nine paired responses span eight local functional
directions, with the integrated Euler density supplying the Lanczos null
vector.

\section{Finite-coefficient operation and response decomposition}
\label{app:finite-dictionary}

This appendix records the deterministic map from the compact cutoff currents
to the finite source observables. The required radial data are
\begin{align}
g_{ij}(\rho,x)={}&g_{(0)ij}+\rho g_{(2)ij}
+\rho^2\left[g_{(4)ij}^{\rm tot}+(\log\rho)h_{(4)ij}\right]
+O(\rho^3\log^p\rho),
\label{eq:app-fg-g}\\
\phi(\rho,x)={}&s\log\rho+\phi_{(0)}+\rho\phi_{(2)}
+\rho^2\left[\phi_{(4)}^{\rm tot}+(\log\rho)\psi_{(4)}\right]
+O(\rho^3\log^p\rho),
\label{eq:app-fg-phi}
\end{align}
with
\begin{equation}
 g_{(4)ij}^{\rm tot}=g_{(4)ij}^{\rm local}+g_{(4)ij}^{\rm state},
 \qquad
 \phi_{(4)}^{\rm tot}=\phi_{(4)}^{\rm local}+\phi_{(4)}^{\rm state}.
 \label{eq:app-total-response}
\end{equation}

\subsection{Solved lower-weight substitution}

The boundary tensors are
\begin{equation}
 X_i=D_i\phi_{(0)},
 \qquad H_{ij}=D_iD_j\phi_{(0)},
 \qquad H=D^2\phi_{(0)}.
 \label{eq:app-boundary-jets}
\end{equation}
and the weight-two data are
\begin{align}
g_{(2)ij}={}&c_{\rm Ric}R_{ij}+c_RRg_{(0)ij}
+c_{XX}X_iX_j+c_{X^2}X_kX^kg_{(0)ij}
+c_HH_{ij}+c_{\Box}Hg_{(0)ij},
\label{eq:app-g2}\\
\phi_{(2)}={}&b_RR+b_{X^2}X_kX^k+b_{\Box}H.
\label{eq:app-phi2}
\end{align}
Define
\begin{align}
U={}&3\ell^2+24\alpha_2s^2-\alpha_0\ell^2s^2-8\alpha_3s^3,
\label{eq:app-u}\\
V={}&6\ell^2+18\alpha_2s^2-\alpha_0\ell^2s^2-8\alpha_3s^3,
\label{eq:app-v}\\
N={}&-3\alpha_3\ell^2+18\alpha_2^2s-24\alpha_2\alpha_3s^2
+\alpha_0\alpha_3\ell^2s^2+8\alpha_3^2s^3,
\label{eq:app-n}\\
C={}&27\alpha_2+3\alpha_0\ell^2-18\alpha_3s
+3\alpha_0\alpha_2s^2-\alpha_0^2\ell^2s^2
-2\alpha_0\alpha_3s^3,
\label{eq:app-c}
\end{align}
\begin{align}
N_{X^2}={}&-54\alpha_2^2\ell^2-9\alpha_0\alpha_2\ell^4
+54\alpha_2\alpha_3\ell^2s-108\alpha_2^3s^2
-18\alpha_0\alpha_2^2\ell^2s^2
-12\alpha_3^2\ell^2s^2
\notag\\
&+3\alpha_0^2\alpha_2\ell^4s^2
+216\alpha_2^2\alpha_3s^3
+6\alpha_0\alpha_2\alpha_3\ell^2s^3
-144\alpha_2\alpha_3^2s^4
+4\alpha_0\alpha_3^2\ell^2s^4+32\alpha_3^3s^5,
\label{eq:app-nq}\\
N_B={}&-18\alpha_2\ell^2-3\alpha_0\ell^4+18\alpha_3\ell^2s
-36\alpha_2^2s^2-6\alpha_0\alpha_2\ell^2s^2
+\alpha_0^2\ell^4s^2
\notag\\
&+48\alpha_2\alpha_3s^3
+2\alpha_0\alpha_3\ell^2s^3-16\alpha_3^2s^4.
\label{eq:app-nb}
\end{align}
The exact coefficient solution is
\begin{align}
c_{\rm Ric}&=-\frac{\ell^2U}{2\Delta_T},
&c_R&=\frac{\ell^2UC-(3\alpha_2-2\alpha_3s)\Delta_T^2}
{12\Delta_T\Delta_S},
\notag\\
c_{XX}&=-\frac{3\alpha_2\ell^2}{\Delta_T},
&c_{X^2}&=-\frac{N_{X^2}}{2\Delta_T\Delta_S},
\notag\\
c_H&=\frac{6\alpha_2\ell^2s}{\Delta_T},
&c_{\Box}&=\frac{2s^2(3\alpha_2-2\alpha_3s)N}
{\Delta_T\Delta_S},
\notag\\
b_R&=\frac{s^2N}{6\Delta_S},
&b_{X^2}&=\frac{N_B}{8s\Delta_S},
&b_{\Box}&=-\frac{sN}{2\Delta_S}.
\label{eq:app-weight-two-dictionary}
\end{align}

\subsection{Counterterm and finite-current coefficients}

The power counterterm density and coefficients are
\begin{equation}
 \mathcal B_{\rm ct}^{(0+2)}=c_0+c_R^{\rm ct}\widehat R+c_X\,X_kX^k,
 \label{eq:app-power-ct}
\end{equation}
\begin{equation}
 c_0=-\frac{2(3+\alpha_0s^2)}{3\ell},
 \qquad
 c_R^{\rm ct}=-\frac{9\ell^2+12\alpha_2s^2-8\alpha_3s^3
 -\alpha_0\ell^2s^2}{6\ell},
 \qquad
 c_X=\frac{3\alpha_2-2\alpha_3s}{\ell}.
 \label{eq:app-power-ct-coefficients}
\end{equation}
Their currents are
\begin{align}
\mathcal J_{\rm ct}^{(0+2)i}{}_j
&=\frac12c_0\delta^i_j-c_R^{\rm ct}\widehat{\mathcal G}^i{}_j
+c_X\left(\frac12X_kX^k\delta^i_j-X^iX_j\right),
\label{eq:app-power-ct-metric}\\
\mathcal J_{\rm ct}^{(0+2)\phi}
&=-2c_X\widehat\Box\phi.
\label{eq:app-power-ct-scalar}
\end{align}

Define the exact finite-coefficient operation
\begin{equation}
 \mathsf F_{\rm SDLH}[F]
 \equiv
 \left.
 \operatorname{Coeff}_{\rho^2(\log\rho)^0}F(\rho)
 \right|_{\substack{g_{(2)}=g_{(2)}^{\star},\,
 \phi_{(2)}=\phi_{(2)}^{\star},\,
 h_{(4)}=h_{(4)}^{\star},\,
 \psi_{(4)}=\psi_{(4)}^{\star}}},
 \label{eq:app-finite-operation}
\end{equation}
where the stars denote the explicit solutions in
Eqs.~\eqref{eq:app-weight-two-dictionary},
\eqref{eq:h4-tf}, \eqref{eq:h4-trace}, and \eqref{eq:psi4}.
The finite canonical currents are
\begin{align}
\mathcal P^i{}_j
&=\mathsf F_{\rm SDLH}
\left[\mathcal J_D{}^i{}_j
+\mathcal J_{\rm ct}^{(0+2)i}{}_j\right],
\label{eq:app-pmetric}\\
\mathcal P_\phi
&=\mathsf F_{\rm SDLH}
\left[\mathcal J_D^\phi
+\mathcal J_{\rm ct}^{(0+2)\phi}\right].
\label{eq:app-pscalar}
\end{align}

The exact normalizable mixing coefficients are
\begin{equation}
 C_g=-\frac{2\Delta_T}{3\ell^3},
 \qquad
 C_\phi=D_g=\frac{8s(3\alpha_2-2\alpha_3s)}{\ell^3},
 \qquad
 D_\phi=\frac{16(\alpha_0\ell^2+6\alpha_2-4\alpha_3s)}{\ell^3}.
 \label{eq:app-response-coefficients}
\end{equation}
They define
\begin{align}
\mathcal P^i{}_j\big|_{\rm resp}
&=C_g\left(g_{(4)}^{{\rm tot},i}{}_j
-\delta^i_j\operatorname{tr}g_{(4)}^{\rm tot}\right)
+C_\phi\phi_{(4)}^{\rm tot}\delta^i_j,
\label{eq:app-metric-response}\\
\mathcal P_\phi\big|_{\rm resp}
&=D_g\operatorname{tr}g_{(4)}^{\rm tot}
+D_\phi\phi_{(4)}^{\rm tot}.
\label{eq:app-scalar-response}
\end{align}
The source-current remainder is therefore
\begin{align}
\mathcal P^i{}_{j,\rm src}
&\equiv\mathcal P^i{}_j-
C_g\left(g_{(4)}^{{\rm tot},i}{}_j
-\delta^i_j\operatorname{tr}g_{(4)}^{\rm tot}\right)
-C_\phi\phi_{(4)}^{\rm tot}\delta^i_j,
\label{eq:app-metric-source-current}\\
\mathcal P_{\phi,\rm src}
&\equiv\mathcal P_\phi-
D_g\operatorname{tr}g_{(4)}^{\rm tot}
-D_\phi\phi_{(4)}^{\rm tot}.
\label{eq:app-scalar-source-current}
\end{align}
Equations~\eqref{eq:app-finite-operation}--
\eqref{eq:app-scalar-source-current} define the covariant source-current
remainder from the compact currents and solved radial data. Equivalent
representatives are related by Bianchi identities, covariant-derivative
commutators, integrations by parts, and the Euler/Lanczos identity.

\subsection{Logarithmic and finite-scheme response vectors}

Write the ordered anomaly coefficient vector as
\begin{equation}
\mathbf a_{\rm reg}=\left(
\frac{2a_E}{3}-\frac{\kappa s^4}{9},
-2a_E,
a_E+a_C,
\frac{\kappa s^2}{3},
0,
-\frac\kappa4,
-\kappa s^2,
-\kappa s,
\frac{2\kappa s^3}{3}
\right),
\label{eq:app-anomaly-vector}
\end{equation}
in the basis \eqref{eq:nine-density-basis}. The cancelling logarithmic
response is
\begin{equation}
 V_{\rm ct}{}^i{}_j=-\sum_{A=1}^9a_{{\rm reg},A}\mathcal V_A{}^i{}_j,
 \qquad
 S_{\rm ct}=-\sum_{A=1}^9a_{{\rm reg},A}\mathcal S_A.
 \label{eq:app-log-response-vector}
\end{equation}
For a finite coefficient vector $\mathbf f=(f_1,\ldots,f_9)$, the complete
one-point functions are
\begin{align}
\langle T^i{}_j\rangle
&=\frac1{8\pi G_5}
\left(\mathcal P^i{}_j-\frac12V_{\rm ct}{}^i{}_j
+\sum_{A=1}^9f_A\mathcal V_A{}^i{}_j\right),
\label{eq:app-stress-tensor}\\
\langle\mathcal O_\phi\rangle
&=\frac1{16\pi G_5}
\left(\mathcal P_\phi-\frac12S_{\rm ct}
+\sum_{A=1}^9f_A\mathcal S_A\right).
\label{eq:app-scalar-onepoint}
\end{align}
At vector weight five, set
\begin{equation}
 \overline{\mathcal E}_{\rho j}^{\rm nonlog}
 =\overline{\mathcal E}_{\rho j}^{[1,0]},
 \qquad
 \overline{\mathcal E}_{\rho j}^{\log}
 =\overline{\mathcal E}_{\rho j}^{[1,1]}.
 \label{eq:app-momentum-coefficients}
\end{equation}
The finite-current construction yields
\begin{equation}
 D_i\langle T^i{}_j\rangle
 -\langle\mathcal O_\phi\rangle D_j\phi_{(0)}
 =\frac{\overline{\mathcal E}_{\rho j}^{\rm nonlog}
 -\tfrac12\overline{\mathcal E}_{\rho j}^{\log}}
 {4\pi G_5\ell}=0,
 \label{eq:app-ward-bridge}
\end{equation}
and
\begin{equation}
 \langle T^i{}_i\rangle-2s\langle\mathcal O_\phi\rangle
 =\frac1{8\pi G_5}\left[-\mathcal A_{\rm reg}
 +\sum_{A=1}^9f_A\left(\operatorname{tr}\mathcal V_A-s\mathcal S_A\right)\right].
 \label{eq:app-trace-bridge}
\end{equation}

\end{document}